\documentclass[letterpaper,twocolumn,10pt]{article}
\usepackage{usenix-2020-09}

\definecolor{myblue}{RGB}{49,133,156}
\definecolor{mygreen}{RGB}{121,149,64}
\definecolor{myred}{RGB}{192,0,0}
\definecolor{mygray}{RGB}{64,64,64}
\definecolor{mygrey}{RGB}{127, 127, 127}
\definecolor{urlblue}{RGB}{22, 31, 201}
\definecolor{refblue}{RGB}{0, 92, 168}
\usepackage{tikz}
\newcommand*\greencircled[1]{\tikz[baseline=(char.base)]{
            \node[shape=circle,draw=white,inner sep=2pt,fill=mygreen,text=white,scale=0.78] (char) {\textbf{#1}};}}
\newcommand*\bluecircled[1]{\tikz[baseline=(char.base)]{
            \node[shape=circle,draw=white,inner sep=2pt,fill=myblue,text=white,scale=0.82] (char) {\textbf{#1}};}}
\newcommand*\redcircled[1]{\tikz[baseline=(char.base)]{
            \node[shape=circle,draw=white,
            inner sep=2pt,
            fill=myred,text=white,font=\bfseries\small,
            text height=5.5pt,
            text depth=1pt,
            scale=0.85] (char) {#1};}}

\usepackage{pifont}
\usepackage{amsmath}
\usepackage{totpages}
\usepackage{comment}
\usepackage{balance}

\usepackage{makecell}

\usepackage{array}

\usepackage{booktabs}

\usepackage[ruled,linesnumbered]{algorithm2e}

\usepackage{multirow}

\usepackage{enumitem}

\usepackage{tcolorbox}

\usepackage{pifont}

\usepackage{cite}
\usepackage{amsmath,amssymb,amsfonts}
\usepackage{algpseudocode}
\usepackage{graphicx}
\usepackage{textcomp}
\usepackage{xcolor}
\def\BibTeX{{\rm B\kern-.05em{\sc i\kern-.025em b}\kern-.08em
    T\kern-.1667em\lower.7ex\hbox{E}\kern-.125emX}}

\usepackage{amsmath}

\usepackage{main_macros}

\begin{document}

\date{}

\title{
Towards Memory-Efficient Traffic Policing in Time-Sensitive Networking
}

\ifarxiv

\author{
    \begin{tabular}{>{\centering}m{3.5cm} >{\centering}m{3.5cm} >{\centering}m{3.5cm} m{3.5cm}}
        {\rm Xuyan Jiang} & {\rm Xiangrui Yang} & {\rm Tongqing Zhou} & {\rm Wenwen Fu} \\[0.5ex]
        {\rm Wei Quan} & {\rm Yihao Jiao} & {\rm Yinhan Sun} & {\rm Zhigang Sun} \\[0.5ex]
        \multicolumn{4}{c}{National University of Defense Technology}
    \end{tabular}
}

\else 

\fi

\maketitle

\begin{abstract}
Time-Sensitive Networking (TSN) is an emerging real-time Ethernet technology that provides deterministic communication for time-critical traffic. At its core, TSN relies on Time-Aware Shaper (TAS) for pre-allocating frames in specific time intervals and Per-Stream Filtering and Policing (PSFP) for mitigating the fatal disturbance of unavoidable frame drift. However, as first identified in this work, PSFP incurs heavy memory consumption during policing, hindering normal switching functionalities.

This work proposes a lightweight policing design called FooDog, which could facilitate sub-microsecond jitter with ultra-low memory consumption. 
FooDog employs a period-wise and stream-wise structure to realize the memory-efficient PSFP without loss of determinism.
Results using commercial FPGAs in typical aerospace scenarios show that FooDog could keep end-to-end time-sensitive traffic jitter <150 nanoseconds in the presence of abnormal traffic, comparable to typical TSN performance without anomalies.
Meanwhile, it consumes merely hundreds of kilobits of memory, reducing >90\% of on-chip memory overheads than unoptimized PSFP design.
\end{abstract}

\section{Introduction}\label{sec: introduction}
Time-Sensitive Networking (TSN)\cite{tsn} is a promising real-time Ethernet technology that builds upon standard Ethernet and has replaced it in many scenarios\cite{introduction}.
Compared to traditional Ethernet, TSN provides \textit{sub-microseconds deterministic jitter} and \textit{lossless forwarding} for time-sensitive periodical traffic while maintaining the best-effort manner for regular traffic.
Such features are crucial to distributed real-time systems such as automotive\cite{jsa,automotive,automotive_2,automotive_3,dg}, aerospace\cite{miura,avionics_1,dp}, and manufacturing\cite{automation,OPC,automation_2,automation_3,automation_4,60802}, and are potentially beneficial for future data center networks and the Internet. 
TSN is becoming popular in real-world implementations of distributed real-time systems. A recent aerospace example is that the Miura 1 suborbital micro launcher adopts TSN as its communication system \cite{miura}.

\begin{figure}[htbp]
    \centering
    \includegraphics[width=\linewidth]{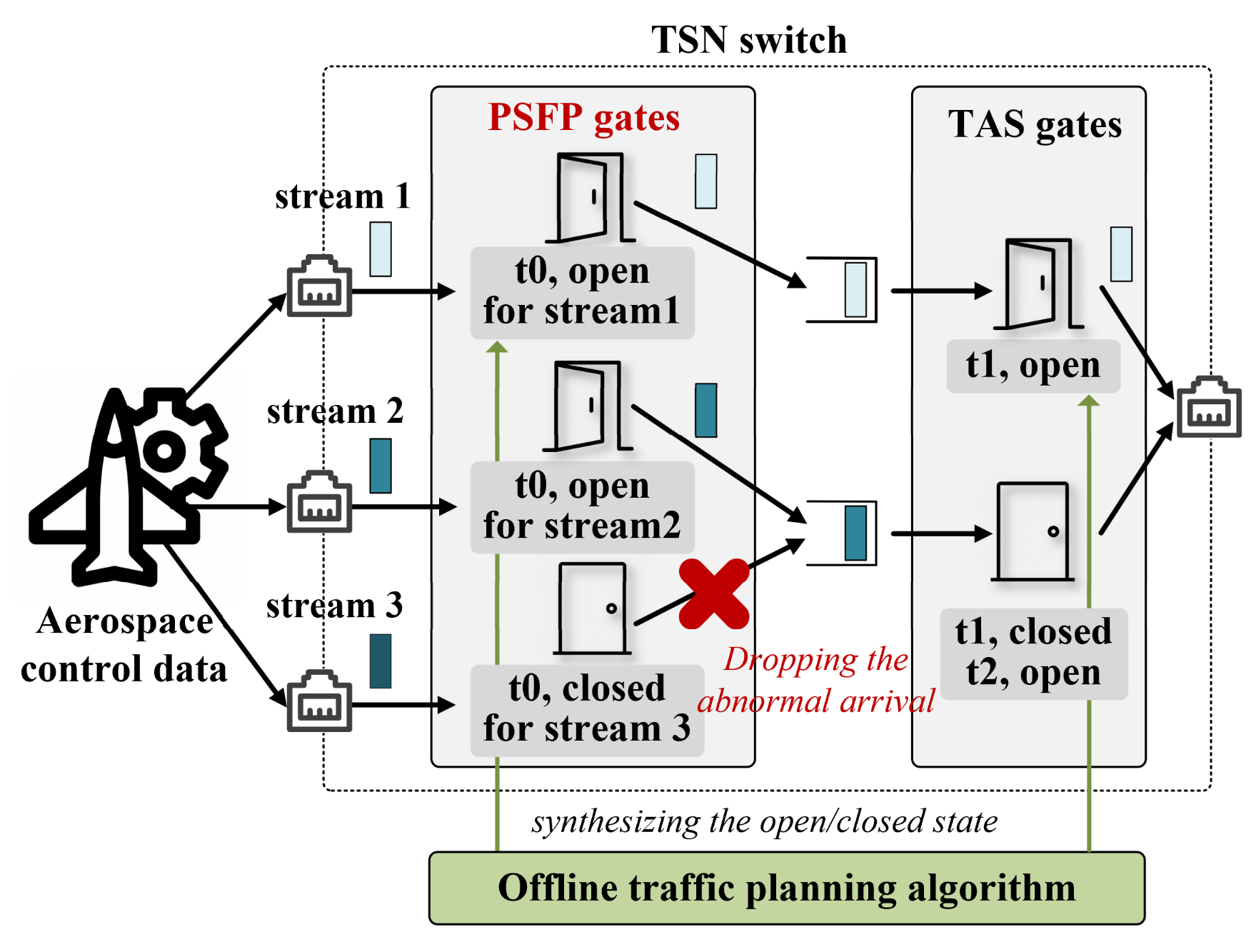}
    \caption{A sketch pipeline of using TAS for deterministic communication and PSFP for abnormality removal.}
    \label{fig: switch architecture}
\end{figure}

The determinism feature of time-sensitive (TS) traffic in TSN is achieved by Time-Aware Shaping (TAS)\cite{Qbv} with the help of global high-precision time synchronization \cite{802.1AS}. TAS ensures transmission with bounded delay and ultra-low jitter.
Each TSN switch utilizes a synchronized timetable called the TAS Gate Control List (GCL). The TAS GCL contains entries that define when and which frames should be forwarded in each network cycle period. By controlling the open/close of the dequeuing gates based on the TAS GCL, the switch can precisely schedule frames on each port.
The core of TAS involves an offline traffic planning algorithm, which synthesizes all the switches' GCLs on a centralized controller in advance \cite{ITP} \cite{dac} \cite{boyang} \cite{array} \cite{RTNS2016}. Traffic planning is a constraints-solving problem \cite{networkplanning} that allocates exclusive time intervals for each frame of each stream\footnote{The traffic is a collection of all the streams in a network, and the stream is a sequence of Ethernet frames.}. This allocation ensures deterministic and contention-free transmission of high-priority and periodic streams.
For example, in Figure~\ref{fig: switch architecture}, streams 1 and 2 are expected and can be successfully scheduled by TAS GCL to arrive at the egress ports at intervals t1 and t2 thanks to TAS gates.

However, in the real world, such an ideal planning would be disturbed by abnormal traffic (e.g., stream 3 in Figure~\ref{fig: switch architecture}), ruining the determinism tenet. In practice, unavoidable manufacturing and environmental factors (e.g., crystal-oscillator drifting\cite{sundial}, synchronization failure\cite{graham}, sensor abnormality\cite{sensor}, space irradiation\cite{emi, see}) may cause streams to abnormally shift away from their scheduled time intervals\cite{error_rate,error_rate_2}, preempting the allocated time intervals of other streams. As reported in\cite{robot_example}, jitters or frame drops in delicate assembling operations on force readings are observed to cause unstable behaviors. To fix this issue, \textit{Per-Stream Filtering and Policing} (PSFP)\cite{Qci} is introduced to filter out abnormal traffic \cite{tsn} by dropping frames that arrive out of the expected arrival time intervals. Similar to TAS, the policing is realized with PSFP GCLs and corresponding hardware gates that record passable states for all streams (e.g., PSFP gates in Figure~\ref{fig: switch architecture}).

This work, for the first time, identifies the heavy memory consumption on-chip introduced by PSFP with the synthesis study of real-world aerospace use cases. Notably, we find that the required memory of PSFP GCLs would, in 25\% of the tested cases, surpass the maximum memory of typical high-end industrial TSN implementation ($\S$\ref{sec: motivation}), hindering table configuration and principal switching functionalities.
Current initial and partial implementations of TSN may overlook this bottleneck, while the future in-depth transformation of Ethernet to TSN, e.g., in aerospace or automotive, may easily approach such memory bounds with much more connected endpoints. 
Via the formulation of standard PSFP design, we further pinpoint that the cause of the heavy memory consumption is that the required memory of PSFP GCLs increases proportionally with the number of streams, the percentage of streams with smaller periods, and the number of ports in a switch ($\S$\ref{sec: rootcause}).

In this context, this work proposes an efficient solution called FooDog\footnote{FooDog is a mythical creature originating from Chinese mythology, tasked with detecting suspicious activities and guarding homes.} to implement PSFP ($\S$\ref{sec: rap design}). Unlike standard PSFP design, in which the planned arrival time windows of streams are all recorded for each frame within the network cycle period, FooDog takes a more efficient approach by only recording the planned arrival window of streams for their first frames within the network cycle period. This compression is made possible by utilizing period-wise GCLs and stream-wise GCLs.
The period-wise GCLs only capture each stream's first appearance during a network cycle period, while the stream-wide GCLs maintain the state (i.e., open/close) of each gate for each stream on the switch data plane. Moreover, such dedicated GCL structures require minimal modifications to the traffic planning algorithms and thus suit well in state-of-the-art TSN ecosystems.

We implement FooDog upon commercial FPGAs ($\S$\ref{sec: implementation}). Experimental results show that, with FooDog, the end-to-end jitters of time-sensitive streams are <150 nanoseconds (i.e., sub-microsecond-level) in the presence of various traffic anomalies, which, we emphasize, are comparable to the performance of typical TSN implementations without abnormal traffic\cite{Fenglin}. More importantly, compared to standard PSFP, FooDog drastically reduces memory usage by $\sim$96\% in use cases of aerospace ($\S$\ref{sec: evaluation}). Such benefits can also be effective for other TSN scenarios with the expansion of TSN in the near future.

\begin{figure*}[htbp]
    \centering
    \includegraphics[width=\linewidth]{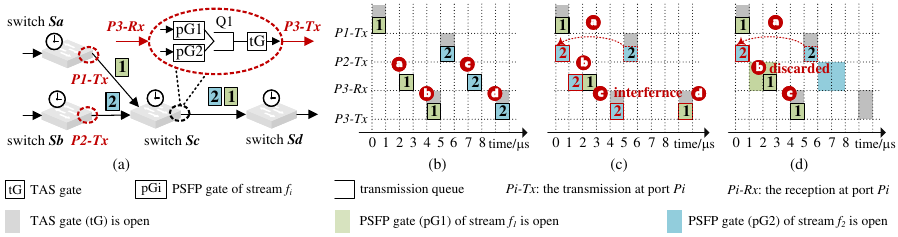}
    \caption{A toy example of the scheduling of TS streams: (a) topology; (b) scheduling of ports when there are no abnormal streams and only TAS gate is enabled; (c) scheduling of ports when there is an abnormal frame $f_2$ and only TAS gate is enabled; (d) scheduling of ports when there is an abnormal frame $f_2$ and both TAS and PSFP gates are enabled.}
    \label{fig: motivation eample}
\end{figure*}

\section{Preliminaries and Related Work}\label{sec: related work}
As TSN uses TAS for deterministic communication and PSFP for abnormal traffic mitigation, 
this section introduces the mechanisms of these two building blocks in the literature.

\subsection{TAS for determinism}
TAS guarantees sub-microsecond-level determinism with the help of the gate control on the data plane and the traffic plan on the control plane. 
Basically, the time-sensitive gate control mechanism executes the dequeue operations for streams according to the gate control list (GCL)\cite{guoqi,Fenglin,gcl_synthesis}, as shown in Figure~\ref{fig: switch architecture}. 
A gate at any given time can be in one of two states: \textit{open} or \textit{closed}. Only if the state of a TAS gate is switched to open could the frames 
be transmitted to the planned transmission queues.

Specifically, the state switching of a gate is pre-planned by the traffic planning algorithm and triggered according to the network-wide synchronized time. 
The traffic planning algorithm calculates a feasible configuration for exclusive time allocations of all TS frames, which is an NP-complete problem\cite{rtss2010}.
The configuration is then synthesized onto the data plane to provide precise controls to the GCLs.

A toy example of the traffic planning of TAS is shown in Figure~\ref{fig: motivation eample}\textcolor{urlblue}{a}. In the example, two streams $f_1$ and $f_2$ pass through four different switches: \textit{Sa}, \textit{Sb}, \textit{Sc}, and \textit{Sd}. The switch \textit{Sc} is at the intersection. Ideally, if there is no abnormal traffic (Figure~\ref{fig: motivation eample}\textcolor{urlblue}{b}), the two streams' enqueue and dequeue operations at \textit{P3} (i.e., the red circle on the top right) would be just as planned. With the help of the traffic planning algorithm, frames of both $f_1$ and $f_2$ will not conflict with each other, although they share the same queue. In the example,
$f_1$ arrives at 2nd $\mu$s (\redcircled{a}) and is scheduled to \textit{Sd} at 4th $\mu$s (\redcircled{b}), while
$f_2$ arrives at 5th $\mu$s (\redcircled{c}) and is scheduled to \textit{Sd} at 9th $\mu$s (\redcircled{d}).

\subsection{Mitigating the impact of abnormal traffic}

However, the design is not perfect with abnormal traffic. Figure~\ref{fig: motivation eample}\textcolor{urlblue}{c} gives an example. When $Sb$ scheduled $f_2$ earlier than expected due to time synchronization errors \cite{sundial,graham} (\redcircled{a}),
$f_2$ may arrive at \textit{Sc} earlier (i.e., 1st $\mu$s) than originally planned. In this case, $f_2$ enters the transmission queue Q1 before $f_1$ (\redcircled{b}). As a result, when the TAS gate $tG$ opens at 4th $\mu$s, $f_2$ instead of $f_1$ is scheduled (\redcircled{c}) first. Such an operation may cause $f_1$ to miss its planned time and lead to the failure of the stream's time-sensitive requirements (\redcircled{d}).

\textbf{PSFP as a remedy.} The scheduling failure demonstrated above can be mitigated by implementing PSFP. Apparently, the root cause of $f_1$'s scheduling failure is that $Sc$ allows an abnormal frame (i.e., $f_2$) to enter Q1 without checking its planned arrival time. 
Thus, by incorporating additional GCLs to track the passage status of each stream, the data plane can detect frames that arrive outside their planned arrival time window. Discarding these abnormal frames can effectively prevent them from disturbing other frames.

Let's revisit the example in Figure~\ref{fig: motivation eample}\textcolor{urlblue}{d}. When PSFP is implemented together with TAS, the PSFP gate $pG2$ at \textit{P3-Rx} could detect that $f_2$ arrives outside its planned arrival time window (6-8th $\mu$s) with the PSFP GCL, so discarding $f_2$ to avoid disturbing downstream switches (\redcircled{b}). For the transmission at \textit{P3-Tx}, since $f_2$ cannot enter the transmission queue Q1, $f_1$ will not be disturbed and retain its determinism as planned (\redcircled{c}).

\begin{figure}[t]
    \centering
    \includegraphics[width=0.33\textwidth]{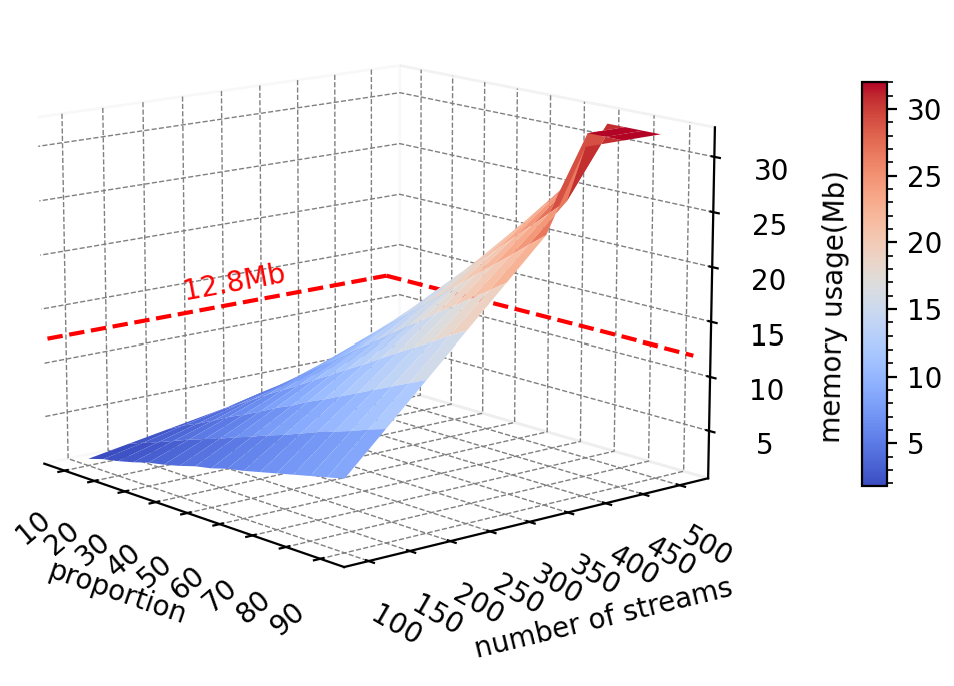}
    \caption{Standard PSFP GCL's BRAM usage of a single TSN switch with four ports. Proportion is the ratio of the number of streams with a period of 1 ms out of the total number of streams.}
    \label{fig: motivation}
\end{figure}

\section{Motivation}\label{sec: motivation}
In this section, we demonstrate our motivation and the necessity for compressing PSFP on switch data plane using a typical TSN use case.

\subsection{Motivating Example}\label{sec: motivation example}

As a matter of fact, the memory consumption of PSFP is not optimal, as revealed by an in-depth analysis of a real-world aerospace use case\cite{avionic_traffic-1}. 
To gain a more intuitive understanding of the practical memory cost of PSFP, we utilize Intel Arria 10 FPGA for a comprehensive memory usage test in real-world use cases. It should be noted that FPGAs are usually preferred over AISCs due to their programmability, as TSN standards are still evolving rapidly. Moreover, recent trends in sectors like aerospace have favored commercially available FPGA switches due to their faster development cycles and mission reusability requirements\cite{nasacots}\cite{ccsds}.
For instance, the Miura-1 suborbital launcher utilized the automotive-grade Z-7030 SoC from the Xilinx Zync 7000 FPGA as its TSN switch\cite{miura}. These COTS devices are constrained by limited FPGA memory capacity (e.g., 9.3Mb\cite{z7030} for Xilinx Zync 7030). Even high-end space-grade FPGA devices such as the Xilinx Radiation Tolerant Kintex UltraScale XQRKU060 only contain 38Mb BRAM\cite{space_grade_fpga} resources.

To test the memory usage in the typical aerospace use case, we modify and implement an open-source TSN switch on Intel Arria 10 FPGA as the TSN switch data plane with standard PSFP integrated.
The FPGA synthesis result is shown in Figure~\ref{fig: motivation}, with the corresponding parameters (i.e., number of streams, stream period, stream composition) aligned with the actual use case\cite{avionic_traffic-1}. Specifically, the number of streams ranges from 100 to 500 with a step of 50. Each tested scale (e.g., 100) consists of streams of length 1ms and length 100ms, with the proportion of 1ms streams ranging from 10\% to 90\%.

As shown in Figure~\ref{fig: motivation}, the on-chip memory usage of PSFP GCLs of a single switch increases linearly with the size and duration of those streams. In particular, we observe that 32.7\% of the tested cases require more BRAM resources than the total memory on the FPGA (the dotted red line in Figure~\ref{fig: motivation}) on a 4-port TSN switch\cite{opentsn}. 
The problem shown in this test is that when the number of streams exceeds 400, which can occur in real-world use cases, the PSFP GCLs' memory usage will exceed the total memory resource on the FPGA (i.e., 32Mb). 
This could lead to insufficient memory resources allocated to the core switch functions such as forwarding table and packet buffering. The example shows that the standard PSFP design, derived directly from the abstraction in the PSFP standard \cite{tsn}, incurs a heavy, if not prohibitive, memory burden on the typical TSN switch data plane.

Future TSN implementation scenarios (e.g., transitioning to the IEEE802.1Q-2022 standard and expanded uses in wireless networks\cite{wsharp} and operational technology\cite{ITOT}\cite{factory}) are expected to generate much larger network traffic, which would be, undoubtedly, accompanied by linearly increased memory usage of PSFP GCLs. One may argue for intuitively using a larger on-chip memory. However, larger network traffic also means more memory resources allocated to L2/L3 forwarding rules, ACLs, NATs, etc. This will further exacerbate the memory resource problem.

\subsection{Root cause analysis}
\label{sec: rootcause}
\begin{figure}[htbp]
    \vspace{-0.1cm}
    \centering
    \includegraphics[width=0.68\linewidth]{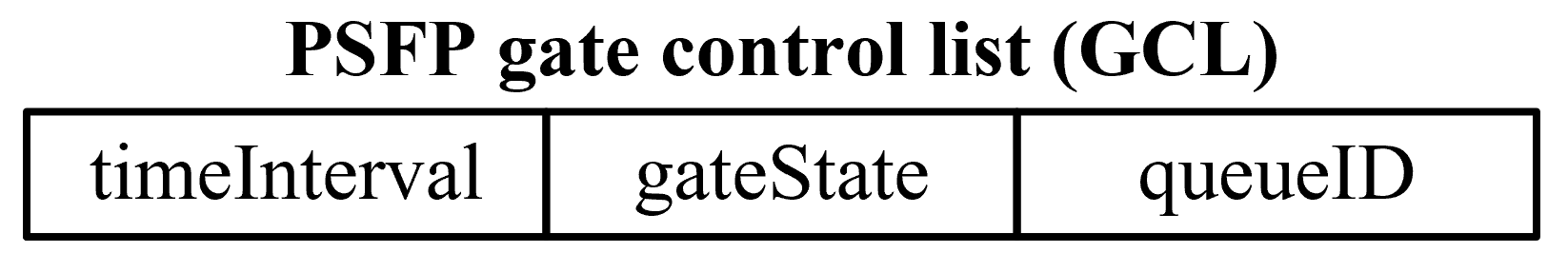}
    \caption{Fields of a PSFP GCL defined in the TSN standard\cite{Qci}}
    \label{fig: gcl structure}
\end{figure}

In this part, we analyze the root cause of heavy memory usage of PSFP in detail. Firstly, let's look at the logical format of the PSFP GCL. As aforementioned, the state of a queue gate is controlled by a GCL.
The fields of PSFP GCL are shown in Figure~\ref{fig: gcl structure}.
PSFP GCL contains three fields: gateState, timeInterval, and queueID. Specifically, the gateState refers to the open/closed states of a gate. The timeInterval represents the interval of the current state.
The queueID denotes the transmission queue that streams will queue up (the queueID is referred to as \textit{itv} in the TSN standard\cite{tsn}, but in this work, we use the term queueID instead for better understanding).

\textbf{Memory usage of standard PSFP GCLs.}
If a device with $P$ ports supports $N$ TS streams, its memory usage of standard PSFP GCLs, denoted as $\mathbb{M}^{std}$, meets Equation~\eqref{eq: naive usage}.
\begin{equation}\label{eq: naive usage}
    \vspace{-0.1cm}
\begin{small}
    \begin{aligned}
        &\mathbb{M}^{std}\geq P\times \sum_{i=1}^{N}M_i^{std} \\
        &M_i^{std} =  W^{std}\times D_i^{std}
    \end{aligned}
    \end{small}
    \vspace{-0.1cm}
\end{equation}

where:

\begin{equation*}
    \vspace{-0.1cm}
    \begin{small}
    \begin{aligned}
        &W^{std} \geq w_{interval}+w_{state}+w_{que} \\
        &D_i^{std} \geq 2\times\frac{T}{T_i}+1
    \end{aligned}
    \end{small}
\end{equation*}

The $M_i^{std}$ is the memory usage of the GCL for the $i$-th TS stream.
The $\sum_{i=1}^{N}M_i^{std}$ represents the total memory usage of $N$ GCLs per port. This is because each of the $N$ streams owns one GCL, as indicated by the name PSFP.
The overall memory usage of the switch should be multiplied by the number of ports $P$.
The $W^{std}$ is the width of the GCL, including timeInterval, gateState, and queueID\cite{Qci}.
The $D_i^{std}$ is the total entries, i.e., the depth of the GCL for the $i$-th TS stream,
which depends on the total number of frames of the $i$-th TS stream within a network cycle period.
The network cycle period refers to the time duration during which all TS streams are scheduled in a recurring manner. 
The length of the cycle period, denoted as $T$, is the least common multiple (LCM) of the periods of all TS streams in the network.
The period of the $i$-th TS stream is denoted as $T_i$.
The $i$-th TS stream with the period $T_i$ has $\frac{T}{T_i}$ frames within a network cycle period.
Each frame requires three entries in the GCL, corresponding to the closed, open, and closed gate states, respectively.
Thus the $i$-th stream with the period $T_i$ needs $2\times\frac{T}{T_j}+1$ GCL entries in total\cite{Qci}.

From Equation~\ref{eq: naive usage} and the preliminary result in Figure~\ref{fig: motivation}, it can be concluded that the memory consumption of naive PSFP GCL is related to the following factors. 

\begin{center}
\begin{tcolorbox}[colback=gray!10,
colframe=black,
width=\linewidth,
arc=1mm,
auto outer arc,
middle=.0cm,
boxrule=0.5pt,
]
Highlight: Memory resource consumption exhibits positive correlations with the number of streams, the number of frames within a network cycle period,
and the number of switch ports.
\end{tcolorbox}
\end{center}

\section{Design of FooDog}\label{sec: rap design}
In this section, we first explore the design space of FooDog, then dive in to the details of FooDog design.

\begin{figure}[htbp]
    \centering
    \includegraphics[width=0.85\linewidth]{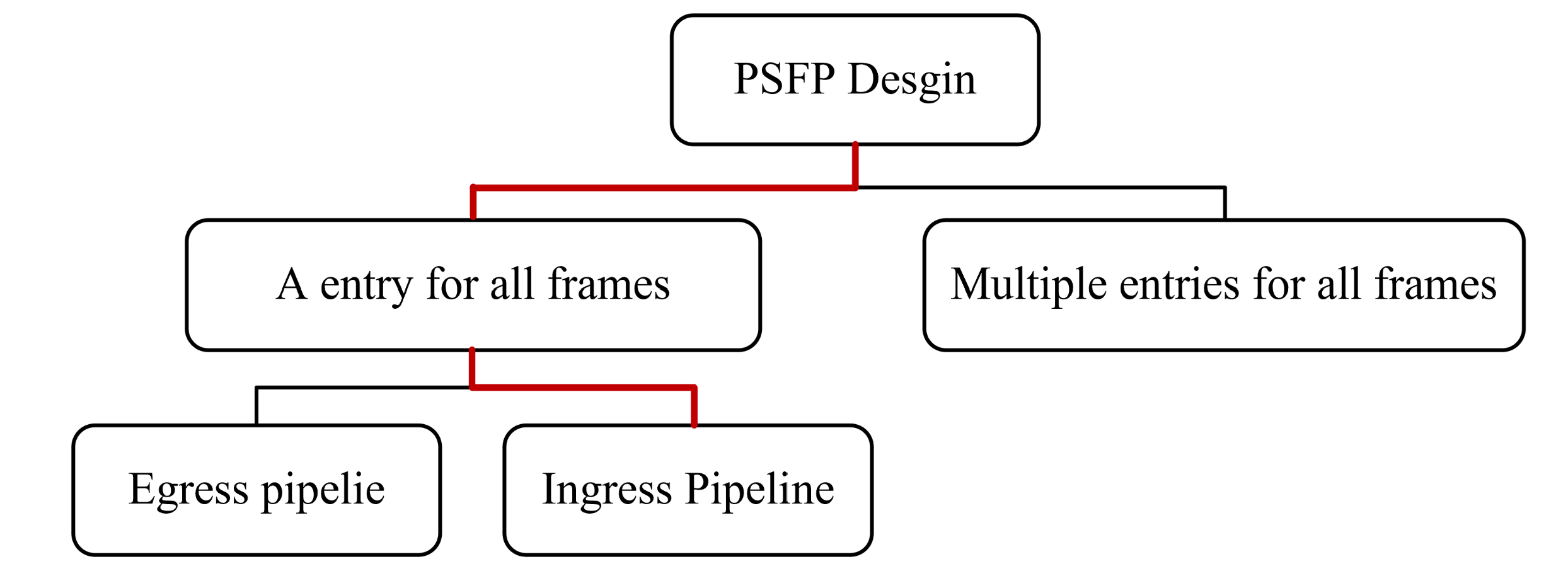}
    \caption{Possible approaches for design PSFP GCL.}
    \label{fig: design space}
\end{figure}

\subsection{Design Decisions}

\textbf{Why a single entry for all frames of a stream?} During a network cycle period, a stream can have multiple frames that are transmitted periodically. However, recording the planned arrival window for each frame within the network cycle period requires a significant number of entries, especially for streams with shorter periods compared to the network cycle period. This observation is supported by experimental results presented in Figure~\ref{fig: memo of FooDog under different port}.

To address this issue, this paper designs a GCL structure that only records the planned arrival time window of the first frame. 
The planned arrival time windows of subsequent frames just need to deviate from the initial window by a certain number of periods.
Subsequent frames within the network cycle period have planned arrival windows that differ from the initial window by a certain number of periods. This approach ensures that the resource consumption of the Per-Stream Filtering and Policing (PSFP) GCL is determined solely by the number of streams and not by the network cycle period.

This design introduces additional constraints, as explained in $\S$\ref{sec: foodog constraint}, which could potentially reduce schedulability and increase end-to-end delay. However, experiments conducted in Section $\S$\ref{sec: solution quality} demonstrate that the impact of these constraints is negligible.

\textbf{Why ingress pipeline?}
According to the standard, PSFP can be implemented in any component (the ingress pipeline, egress pipeline, etc.) of the switch data plane before the TAS scheduling gates. However, the ingress pipeline connects to the ingress port, and the egress pipeline connects to the egress port.
In this work, PSFP is deployed in the ingress pipeline due to its proximity to the upstream devices. By enforcing policing at the ingress pipeline, abnormal streams can be effectively prevented from interfering with the downstream device processing.

When deployed in the ingress pipeline, the memory consumption of PSFP GCL exhibits a linear relationship with the number of ports, which is the same as the standard PSFP design (as depicted in Equation~\ref{eq: naive usage}). However, thanks to the proposed memory-efficient GCL structure, the overall resource utilization within a switch remains acceptable. This observation is further supported by the experimental results presented in $\S$\ref{sec: memory}.

\subsection{Overview}
\begin{figure}[htbp]
\centering
\includegraphics[width=\linewidth]{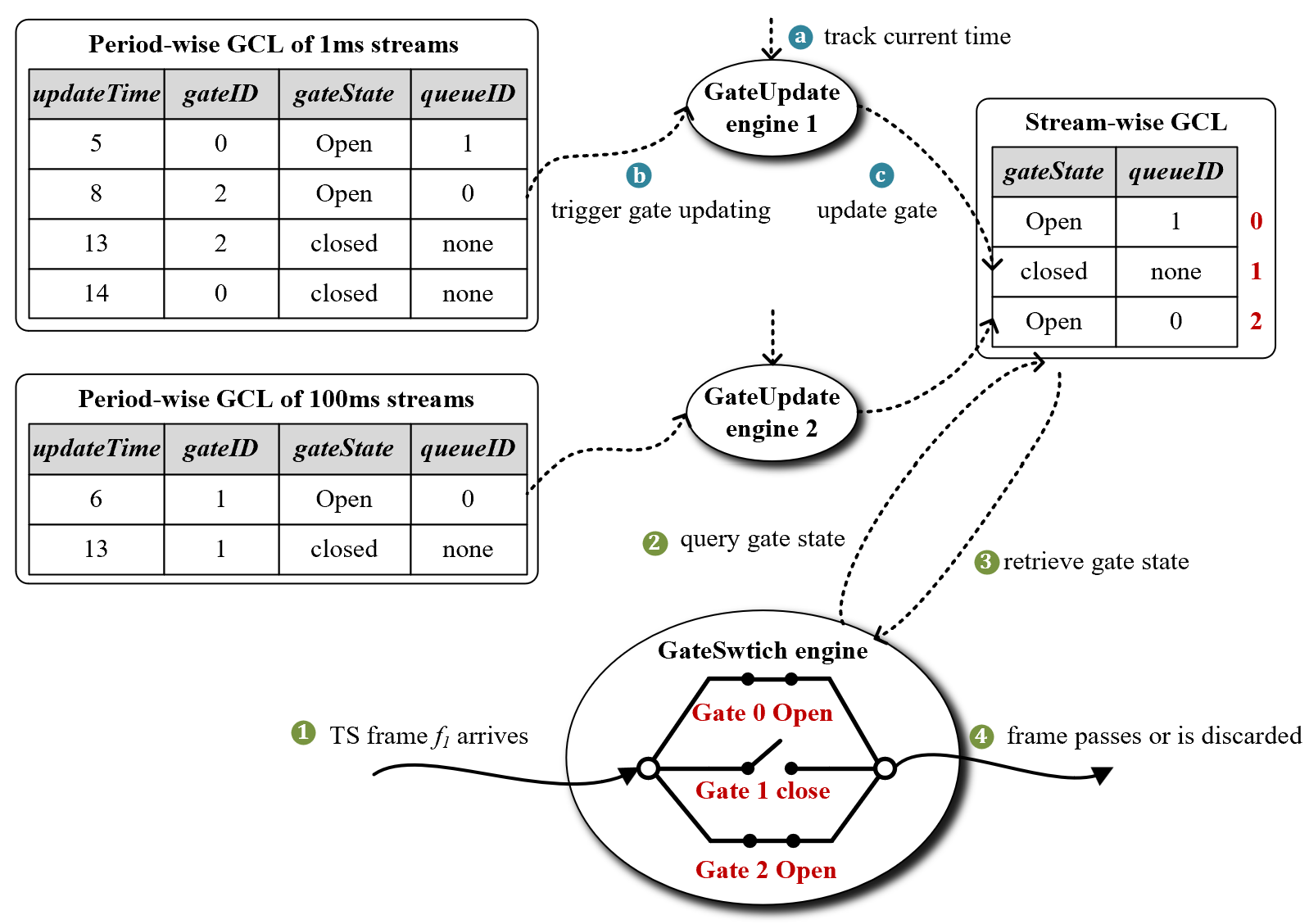}
\caption{FooDog system diagram
}
\label{fig: overview}
\end{figure}

FooDog aims to maintain ultra-low memory usage while keeping sub-microsecond-level determinism.
FooDog introduces Period-wise GCLs to capture the planned arrival time windows of the first streams and a Stream-wise GCL to capture the open/closed status of gates at the current time.
The Period-wise GCL is crucial for achieving memory efficiency in the FooDog system. It only stores information about the planned arrival time window of the first frame within the network cycle period. This means that the amount of resources used by FooDog is not dependent on the number of frames within the network cycle period.
Besides, FooDog also incorporates policing engines to maintain the gates' states and enforce traffic policing.
The system diagram of the FooDog architecture can be illustrated in Figure~\ref{fig: overview}.

\subsection{Period-wise and Stream-wise GCL}

Period-wise GCLs and Stream-wise GCLs are interconnected to perform PSFP functions and serve different purposes.
The Period-wise GCLs schedule streams according to different periods and record their planned arrival time window, while the Stream-wise GCL captures the current open or closed status of all gates of all streams.

\textbf{Period-wise GCLs} records the open and closed timings of gates to capture streams' planned arrival time windows.
Period-wise, GCLs are organized according to the stream's period.
There are multiple Period-wise GCLs based on different stream periods. 
Each Period-wise GCL captures the planned arrival window of streams with the same periods.

Each Period-wise GCL contains four fields. 
The \textit{updateTime} indicates the time when the entry should be updated into the Stream-wise GCL. 
The \textit{gateState} indicates the open or closed state. 
The \textit{queueID} specifies the transmission queue to which the stream is directed. 
The \textit{gateID} specifies which gate the entry updates, as well as the actual address in the Stream-wise GCL where the entry should be written.
The entries in each Period-wise GCL are sorted in ascending order based on their \textit{updateTime}, indicating the chronological order of updates.

The Period-wise GCL introduces \textit{pGCL cycle period}.
The pGCL cycle period is the time duration required for the Period-wise GCL to execute from the first entry to the last entry.
Each Period-wise GCL has its own pGCL cycle period.
The length of a pGCL cycle period is equal to the period of the streams whose planned arrival time windows are captured by the Period-wise GCL.
The length of the pGCL cycle period of the $j$-th Period-wise GCL is denoted as $pT_j$.
For example, a Period-wise GCL captures the planned arrival time windows of streams with a period of 1 ms; then its pGCL cycle period is 1 ms.
The length of a pGCL cycle period is different from the network cycle period, whose length is defined as the LCM of all streams' periods in the network.
Thus, a Period-wise GCL may be executed multiple times in a recurring manner within a network cycle period.

Since the GPL cycle period equals the streaming period, the Period-wise GCL could use only two entries to capture a stream's opening and closing timing. 
By executing the $j$-th Period-wise GCL $\frac{T}{pT_j}$ times, we could capture all the planned arrival windows of streams within the entire network cycle period.

The \textbf{Stream-wise GCL} captures the open or closed status of all gates at the current time. The Stream-wise GCL consists of two fields. The \textit{gateState} indicates the $idx$-th gate is in the open or closed state, where $idx$ is the index of this entry in the Stream-wise GCL. The \textit{queueID} identifies the transmission queue to which the frames of the stream will be directed.
The content of Period-wise GCL is updated into the Stream-wise GCL when a frame arrives at FooDog (see workflow in $\S$\ref{workflow}).

\textbf{Memory usage of FooDog.} If a device with P ports support $N$ TS streams, its memory usage of FooDog, denoted as $\mathbb{M}^{FooDog}$, meets Equation~\ref{eq: FooDog usage}. 

\begin{equation}\label{eq: FooDog usage}
\begin{small}
\begin{aligned}
        \mathbb{M}^{FooDog}\geq P\times \left(W^{pGCL}\times D^{pGCL}+W^{sGCL}\times D^{sGCL}\right)
\end{aligned}
\end{small}
\vspace{-0.1cm}
\end{equation}

where:

\begin{equation*}
\begin{small}
\begin{aligned}
    &W^{pGCL}\geq w_{time}+w_{gate}+w_{state}+w_{que},w_{gate}\geq \lceil \log_{2}N\rceil\\
    &W^{sGCL}\geq w_{state}+w_{que}\\
    &D^{pGCL}\geq 2\times N, D^{sGCL}\geq N
\end{aligned}
\end{small}
\vspace{-0.1cm}
\end{equation*}

The $W^{pGCL}$ is the width of a Period-wise GCL including updateTime, gateID, gateState and queueID.
The $W^{sGCL}$ is the width of a Stream-wise GCL including gateState and queueID.
The width of the gateID fields in the Period-wise GCL should not be less than $\lceil \log_{2}N\rceil$, which is enough to represent $N$ gates for $N$ streams.
The $D^{pGCL}$ is the \textit{total} depth of all Period-wise GCLs.
Since each stream requires two entries to record its planned arrival time windows, the total entries occupied by $N$ streams is $2\times N$.
The $D^{sGCL}$ is the depth of a Stream-wise GCL, which should not be less than $N$ in order to capture $N$ gates' status for $N$ streams.

\begin{center}
\begin{tcolorbox}[colback=gray!10, 
colframe=black, 
width=\linewidth, 
arc=1mm,
auto outer arc,
middle=.0cm,
boxrule=0.5pt,
]
Highlight: The resource consumption of FooDog is independent of the number of frames within the network cycle period and only exhibits positive correlations with the number of streams and the number of ports.
\end{tcolorbox}
\end{center}

\subsection{Policing Engines}
FooDog introduces two types of policing engines to maintain the gates' states and enforce traffic policing: the GateUpdate engines and the GateSwitch engine.

The \textbf{GateUpdate engines} maintain the current gate statuses by updating the contents of the Period-wise GCL into the Stream-wise GCL.
Each GateUpdate engine executes a single Period-wise GCL. There are multiple GateUpdate engines in the FooDog. The number of GateUpdate engines equals the number of streams supported by the switch with different periods.
For example, the FooDog in Figure~\ref{fig: overview} has two GateUpdate engines because streams have two different periods, i.e., 1ms and 100ms.

The update operation of GateUpdate engines follows a predetermined time-triggered pattern, as indicated by the \textit{updateTime} fields in the Period-wise GCLs.
A GateUpdate engine sequentially retrieves the entries from the Period-wise GCL, driven by the global synchronized time.
When the global time is updated, the engine compares whether the current entry's update time has expired. If it has, the engine updates the \textit{gateState} and \textit{queueID} into the Stream-wise GCL.
The address of the entry in the Steam-wise GCL is indicated by the \textit{gateID} of the entry. 
After the engine finishes executing all entries in the Period-wise GCL, it loops back to the first entry and starts over again. The time it takes to execute the entire Period-wise GCL is the length of the pGCL cycle period.

Though the Period-wise GCL only records the planned arrival time window of the first frame within the network cycle period, by repeatedly executing the Period-wise GCL with a period of pGCL cycle period, the GateUpdate engine could express the planned arrival time windows of all frames within the network cycle period.

The entries in each Period-wise GCL are arranged in ascending order according to their \textit{updateTime}. As a result, the time complexity of the update operations performed by the GateUpdate engines is \verb|O(1)| since they are driven by time. Note that there are no conflicts between the update operations of different GateUpdate engines because each engine is responsible for maintaining different gates.

The \textbf{GateSwitch engine} enforces the traffic policing according to the \textit{gateStates} in the Stream-wise GCL.
When a frame is received, the GateSwitch engine extracts the \textit{streamID} from it, which is then used as the address to query the Stream-wise GCL. This is possible because each stream in FooDog has its own dedicated gate, and the address of a gate in the Stream-wise GCL equals its index.
If the \textit{gateState} in the returned entry of the Stream-wise GCL is open, the frame is directed to the transmission queue specified by the \textit{queueID}. Otherwise, the frame is discarded.
The time complexity of the lookup operations performed by the GateSwitch engine is \verb|O(1)| because the \textit{streamID} is used as the address of the Stream-wise GCL.

\subsection{Workflow}\label{workflow}
To provide a comprehensive understanding of FooDog, the workflow of the update plane and police plane are described as follows.

\textit{In the update plane, the GateUpdate engines work with Period-GCLs and the Stream-wise GCL to maintain the gate status.}
Take the GateUpdate engine 1 in Figure~\ref{fig: overview} as an example.
A GateUpdate engine maintains a current address pointer.
When the global synchronized time updates (\bluecircled{a}), the GateUpdate engine retrieves the entry pointed by the current address pointer and compares the \textit{updateTime} of the entry with the current time (\bluecircled{b}).
If the \textit{updateTime} expires, the GateUpdate engine updates the \textit{gateState} and \textit{queueID} into the Stream-wise GCL (\bluecircled{c}).
The \textit{gateID} is used as the address.
Then, the GateUpdate engine advances the address pointer to the next entry.

\textit{In the police plane, the GateSwitch engine works with the Stream-wise GCL to enforce traffic policing.}
When a TS frame arrives at the GateSwitch engine (\greencircled{1}), the GateSwitch engine extracts the \textit{streamID} from the frame.
Then, the streamID is used as the address to query the Stream-wise GCL(\greencircled{2}).
The GateSwitch engine retrieves the \textit{gateState} and \textit{queueID}(\greencircled{3}).
If the \textit{gateState} is closed, the frame is discarded.
If the gateState is open, the frame is directed into the transmission queue indicated by the \textit{queueID} (\greencircled{4}).

\subsection{FooDog Constraints}\label{sec: foodog constraint}
In FooDog, the Period-wise GCL solely captures the planned arrival window of the first frame of a stream during the network cycle period.
This introduces two constraints, referred to as the FooDog constraints, into the traffic planning algorithm.
The first constraint is that all frames within a network cycle period must pass through the same transmission queue at the same hop along the routing path.
The second constraint is that the transmission times of different frames belonging to the same stream are strictly periodic.
The FooDog constraints are not new, which has been implemented\cite{RTNS2016}. Therefore, we directly adopted the algorithm\cite{RTNS2016} for FooDog.
For clarity, we refer to the adopted algorithm as the FooDog algorithm, and the algorithm without the FooDog constraints is called the Comp algorithm.

Theoretical considerations suggest that incorporating the FooDog constraints into traffic planning problems may limit the available solution space. However, our experimental results (Figure~\ref{ex-fig: schedulablity}) demonstrate that this reduction is negligible. The TS traffic occupies a small portion of the overall bandwidth compared to the link bandwidth (Figure~\ref{ex-fig: bandwidth}), indicating that the solution space remains sufficiently large. Additionally, there is a clear trend showing that the bandwidth used by TS flows is relatively small when compared to TSN's continuously increasing link bandwidth that exceeds 25Gbps\cite{60802,dp,dg,53570}. Therefore, implementing the FooDog constraints would not result in significant compromises in high-bandwidth scenarios.

Moreover, the configuration of the Period-wise GCLs in the FooDog is derived from the results of a planning algorithm. The mapping from the upstream device's scheduled time to the downstream device's planned arrival time window is straightforward and explained in detail in Appendix~\ref{sec: generation of Period-wise GCL}.

\subsection{Discussions}

We note that although the proposed FooDog design can mitigate interference from abnormal streams, it cannot eliminate interference caused by errors (e.g., crystal-oscillator drifting\cite{sundial}, synchronization failure\cite{graham}, sensor abnormality\cite{sensor}, space irradiation\cite{emi, see}), which may result in abnormal streams. This is because FooDog drops the abnormal streams indistinguishably, while the abnormal streams should still be transmitted deterministically to their destination under normal circumstances. Eliminating error interference requires additional reliability mechanisms like frame replication\cite{802.1CB}, yet not in the scope of this work.

\begin{figure*}[t]
    \centering
    \includegraphics[width=\linewidth]{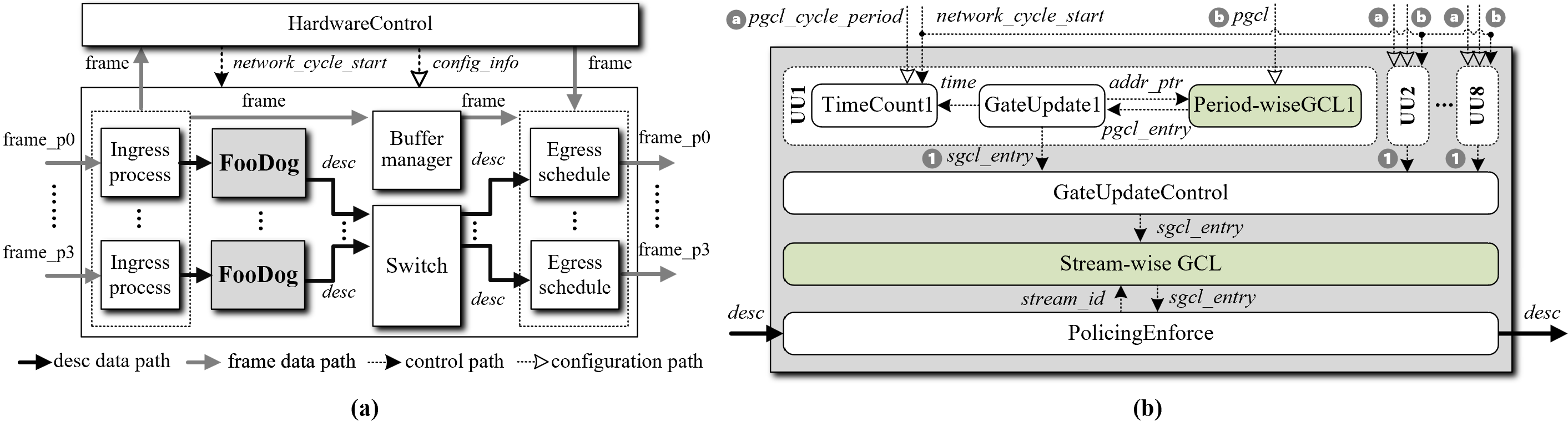}
    \caption{Implementation diagram of FooDog prototype: (a) architecture of the TSN switch; (b) FooDog implementation.}
    \label{fig: implementation}
\end{figure*}

\section{Implementation}
\label{sec: implementation}
Our implementation consists of the FooDog prototype on the FPGA hardware, as well as the FooDog algorithm and the Comp algorithm\footnote{Source code of the implementation will be made available for the community.}.

\textbf{FooDog hardware.}
Figure~\ref{fig: implementation}\textcolor{urlblue}{(a)} illustrates the implementation of the switch data plane and the location of the FooDog prototype. The FooDog prototype is situated between the ingress process module and the switch module. The switch data plane processes the frame descriptor, which includes the metadata of a frame. The frame descriptor comprises 64 bits, with three fields associated with the FooDog, namely \textit{queueID}, \textit{streamID}, and \textit{discard}. The \textit{queueID} field contains 3 bits, indicating the queue that the frame enters. The \textit{streamID} field contains 14 bits, indicating the stream to which the frame belongs. This design enables support for up to 16384 streams, which is sufficient for TSN-targeted applications. The \textit{discard} field contains 1 bit, indicating whether or not the frame is discarded.

The FooDog prototype supports 500 streams with eight arbitrary periods, regardless of their network cycle period length. The widths of each field in the Period-wise GCL and Stream-wise GCL are as follows: the \textit{updateTime} is 32 bits, in accordance with the TSN standard specification\cite{Qci}; the \textit{gateID} is 9 bits, which is adequate to represent the \textit{gateID} of 500 streams; the \textit{gateState} is 1 bit; and the \textit{queueID} is 3 bits, as there are eight transmission queues per port.

To support arbitrary percentages of streams for eight periods, we employ eight Period-wise GCLs with a length of 1000. One Period-wise GCL is sufficient to capture the planned arrival time window for all 500 streams. The depth of the Stream-wise GCLs is equal to the number of streams, which is 500. Both the Period-wise GCL and Stream-wise GCL are implemented using dual-port RAM. The required block RARM resources are presented in the experimental section ($\S$\ref{sec: determinism}).

The implementation of FooDog is illustrated in Figure~\ref{fig: implementation}\textcolor{urlblue}{(b)}, and its logical resource usage is demonstrated in $\S$\ref{sec: determinism}. FooDog comprises four types of modules: 8 GateUpdate modules associated with 8 Period-wise GCLs, TimeCount modules for generating time for each GateUpdate module, a GateUpdateControl module, and a PolicingEnforce module. An individual GateUpdate module, TimeCount module, and the RAM of a Period-wise GCL together form an UpdateUnit (UU). The UU serves as the fundamental unit for maintaining the gate state of all frames sharing the same cycle period. In the FooDog prototype, there are 8 UUs. The functionalities of each module are explained below.

The TimeCount module takes the \textit{network\_cycle\_start} signal received from the HardwareControl module in Figure~\ref{fig: implementation}\textcolor{urlblue}{(a)} and breaks it down into separate time based on the \textit{pgcl\_cycle}. The \textit{network\_cycle\_start} signal indicates the start of a network cycle period.
The \textit{pgcl\_cycle} indicates the length of the pGCL cycle period.

The GateUpdate module maintains the current address pointer of the ongoing entry \textit{addr\_ptr} in the RAM of the Period-wise GCL. When the TimeCount module updates the \textit{time}, the GateUpdate module compares the \textit{updateTime} field of the ongoing entry with the current \textit{time}. If the \textit{updateTime} is greater than the current \textit{time}, the GateUpdate module notifies the GateUpdateControl module about the \textit{sgcl\_entry} including \textit{gateID}, \textit{gateState}, and \textit{queueID}. Additionally, it increments the address pointer \textit{addr\_ptr}.

The GateUpdateControl module sequentially writes the updated \textit{sgcl\_entry} from the UpdateEngine into the Stream-wise GCL, thereby avoiding any potential write conflicts that may occur between multiple UUs.

The PolicingEnforce module assesses the eligibility of frames by referencing the RAM of the Stream-wise GCL. This module extracts the \textit{streamID} from the descriptor \textit{desc} and uses it as an address to access Stream-wise GCL RAM. Based on the \textit{gateState} of the retrieved \textit{sgcl\_entry}, the PolicingEnforce module determines whether the frame is eligible. If the \textit{gateState} field is 0, the \textit{discard} of the descriptor is set to 1. Conversely, if the \textit{gateState} field is 1, the \textit{discard} of the descriptor is set to 0. Additionally, the \textit{queueID} of the descriptor is filled with the content of the \textit{queueID} from the \textit{sgcl\_entry}.

FooDog relies on the HardwareControl module in Figure~\ref{fig: implementation}\textcolor{urlblue}{(a)} to provide configuration support. There are two types of configuration for the FooDog prototype. The first type involves configuring all the Period-wise GCLs \textit{pgcl}. The second category involves configuring the length of the \textit{pgcl\_cycle} for the TimeCount module. These configurations are calculated by a centralized network controller that runs planning algorithms. 
The configuration protocol is Netconf\cite{netconf}.

\textbf{Software for FooDog.}
The FooDog algorithm and the Comp algorithm are implemented using Python programming language. The constraints of the FooDog algorithm are stated in the paper\cite{RTNS2016}, while the constraints of the Comp algorithm are described in Appendix~\ref{appendix: constraints}. An open-source SMT solver called z3py (version 4.8.15)\cite{z3py} is used to solve these constraints. We input all the constraints into the SMT solver and wait for it to find feasible solutions for all unknown variables. The reason we use an SMT solver instead of other approaches, such as heuristics, is that SMT solvers are mature and inherently stable. Our goal is not to optimize the algorithms. Instead, we aim to conduct comprehensive experiments to compare the solution quality between algorithms with and without the FooDog constraint described in $\S$\ref{sec: foodog constraint}.

\textbf{Others.}
The open-source TSN project includes a centralized controller responsible for configuring the TSN switches. To support the configuration of FooDog, the controller has been improved by integrating the FooDog configuration, which is programmed in the C programming language.

\section{Evaluation}\label{sec: evaluation}

We evaluate FooDog by answering the following questions:

1. How much memory does FooDog consume in typical use cases? 
    ($\S$\ref{sec: memory})

2. Does FooDog guarantee sub-microsecond-level determinism against abnormal traffic? ($\S$\ref{sec: determinism})

3. How about the solution quality of the FooDog algorithm in terms of runtime and other relevant metrics? ($\S$\ref{sec: solution quality}
)

\begin{figure}[htbp]
        \centering
            \subfloat[Aerospace topology]{
            \includegraphics[width=0.6257\linewidth]{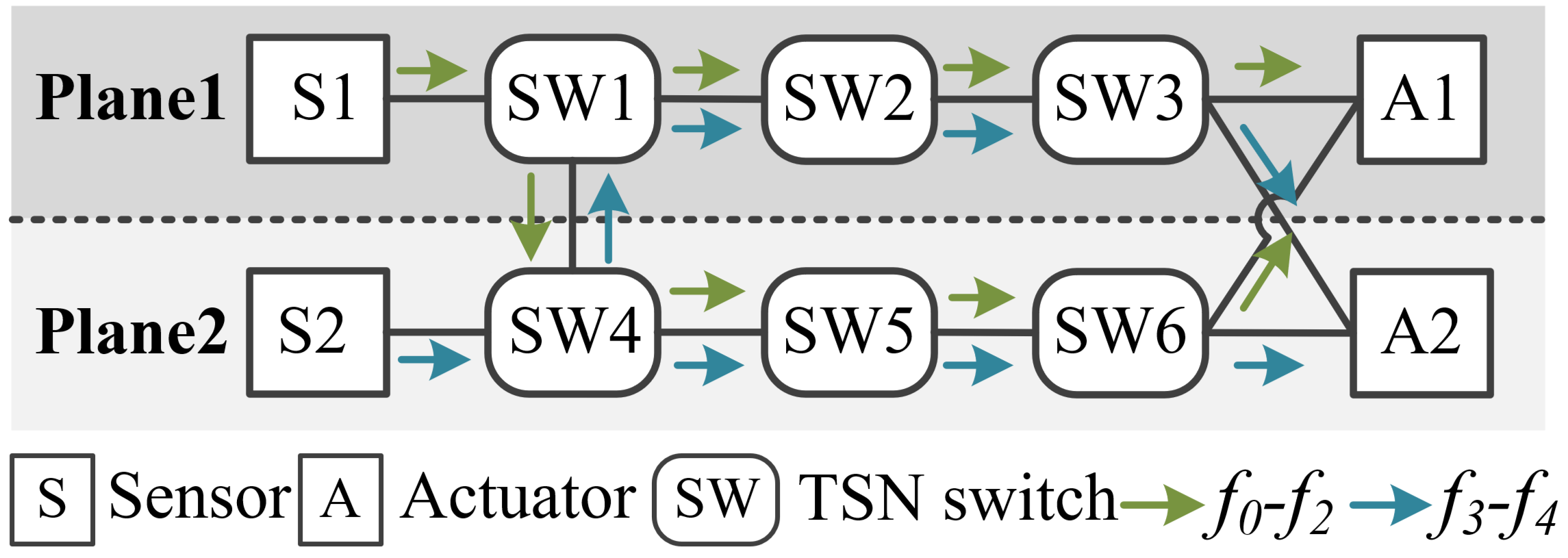}
            \label{fig: avionic topo}}
        \centering
            \subfloat[Aerospace demo]{
            \includegraphics[width=0.2846\linewidth]{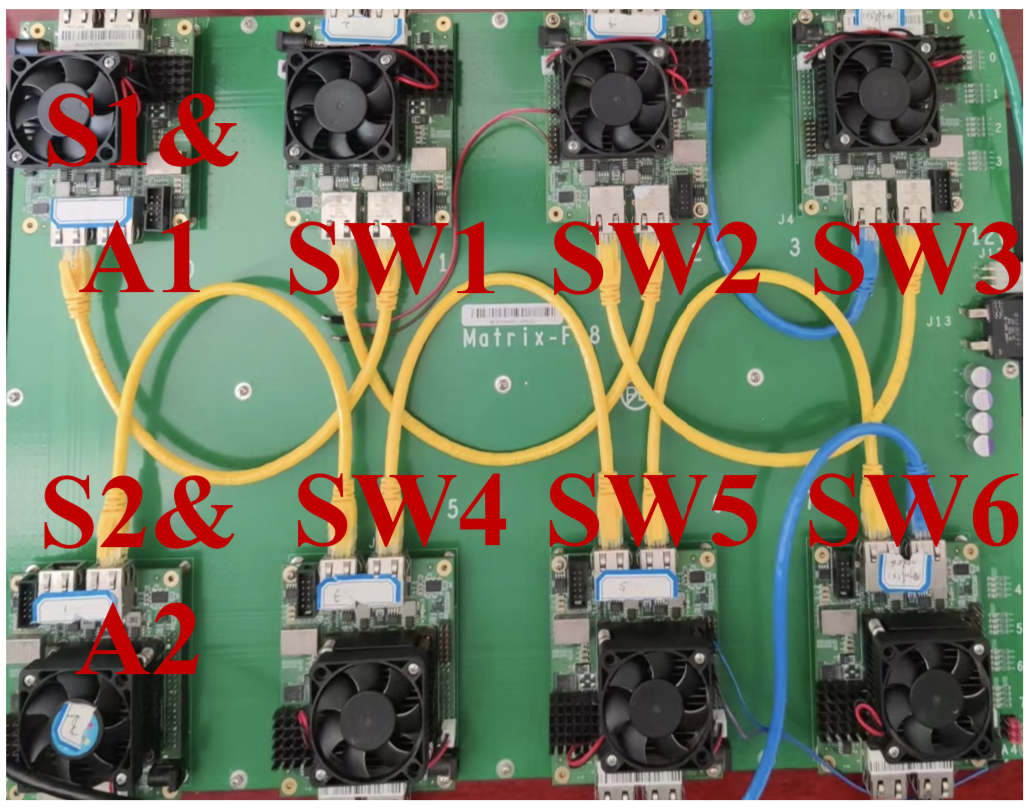}
            \label{fig: avionic real-world topo}}
            \hfill
    \caption{Aerospace network.}
    \label{fig: aerospace}
\end{figure}

\textbf{Environmental Setup.}
To evaluate the FooDog prototype, an experimental scenario based on a spacecraft\cite{avionic_topo} is adopted as shown in Figure~\ref{fig: aerospace}. 
There are two sensors $S1-S2$, two actuators $A1-A2$, and six TSN switches $SW1-SW6$.
The topology consists of two parallel transmitting planes for redundancy. We focus on the transmission at Plane 1.

The FooDog algorithm and the Comp algorithm ran on an Intel Core i7-8550U 64-bit CPU @ 1.80GHz with 128 GB of RAM. 
The FooDog algorithm and the Comp algorithm are evaluated on various topologies, including NASA's Orion Crew Exploration Vehicle topology\cite{nasa}, an automotive topology\cite{auto}, and a train topology\cite{train}.
Since there is no available stream dataset for TSN\cite{tcad}, we generate our stream randomly.
The streams are with a period of either 1ms or 100ms, with a frame length of 100Bytes. The end-to-end latency requirement was set at 100 $\mu$s or 1000 $\mu$s, while the jitter requirement was set at 10 $\mu$s or 100 $\mu$s.
The maximum hop count for TS streams is set to be 7.

For the scenarios above, the link speed is 1000 Mbps. 
The synchronization precision at all links is under 48 ns. 
The link delay ranges from 0.3 $\mu$s to 1.2 $\mu$s.
Each egress port is equipped with two transmission queues for TS streams.
Unless otherwise specified, the number of switch ports is 4.

\begin{figure*}[htbp]
  \centering
  \begin{minipage}{0.318\textwidth}
    \includegraphics[width=\linewidth]{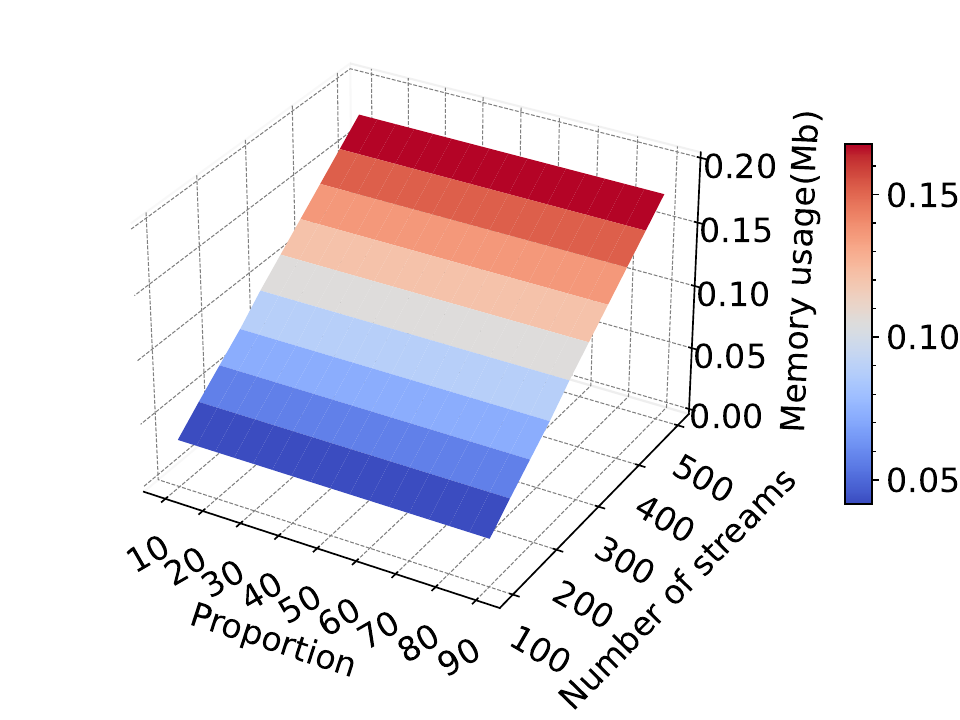}
    \caption{FooDog's theoretical memory usage}
    \label{fig: theo memo of FooDog}
  \end{minipage}\hfill
  \begin{minipage}{0.313\textwidth}
    \includegraphics[width=\linewidth]{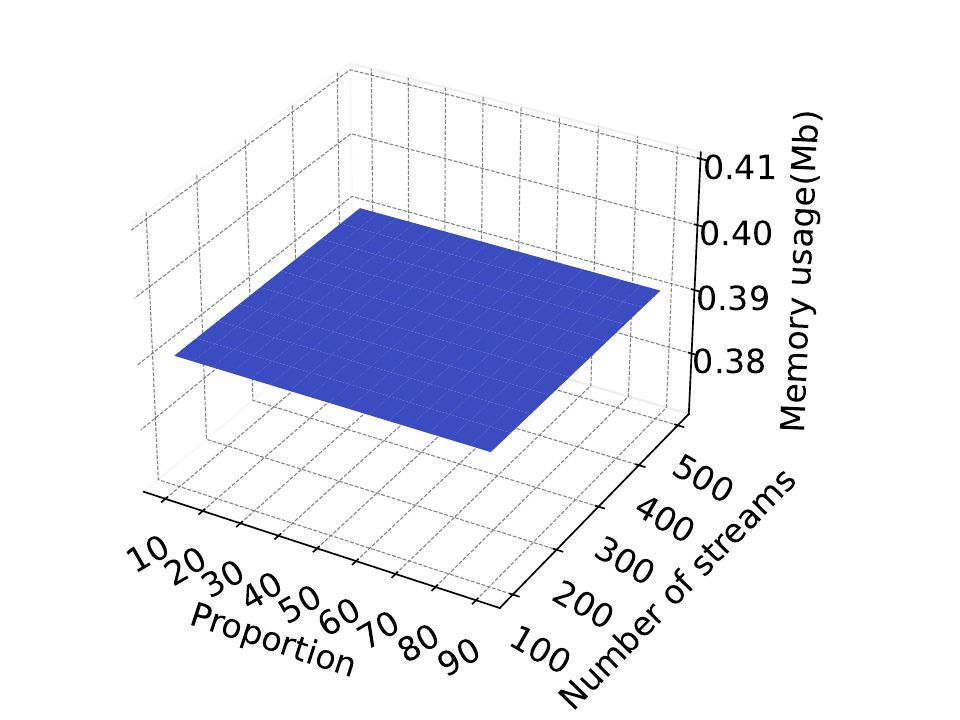}
    \caption{FooDog's actual memory usage}
    \label{fig: exp memo of FooDog}
  \end{minipage}\hfill
  \begin{minipage}{0.358\textwidth}
    \includegraphics[width=\linewidth]{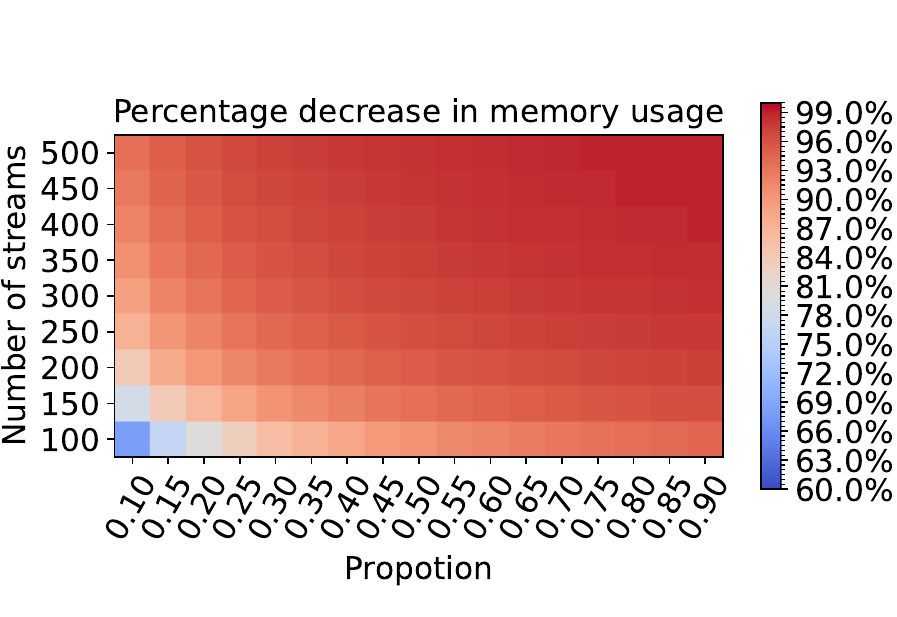}
    \caption{Heatmap of reduced memory usage}
    \label{fig: percentage decrease}
  \end{minipage}\hfill

\end{figure*}

\begin{figure}[!htbp]
    \centering
    \begin{minipage}[b]{0.45\linewidth}
        \includegraphics[width=\linewidth]{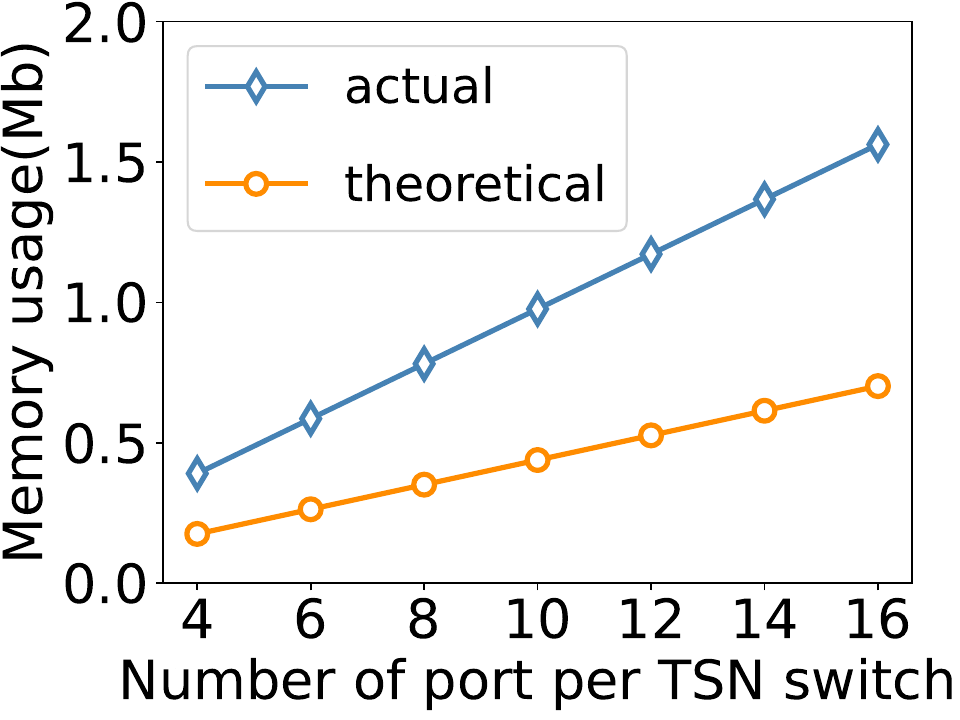}
        \caption{Resource consumption at different numbers of ports}
        \label{fig: memo of FooDog under different port}
    \end{minipage}
    \hspace{0.05\linewidth}
    \begin{minipage}[b]{0.45\linewidth}
        \includegraphics[width=\linewidth]{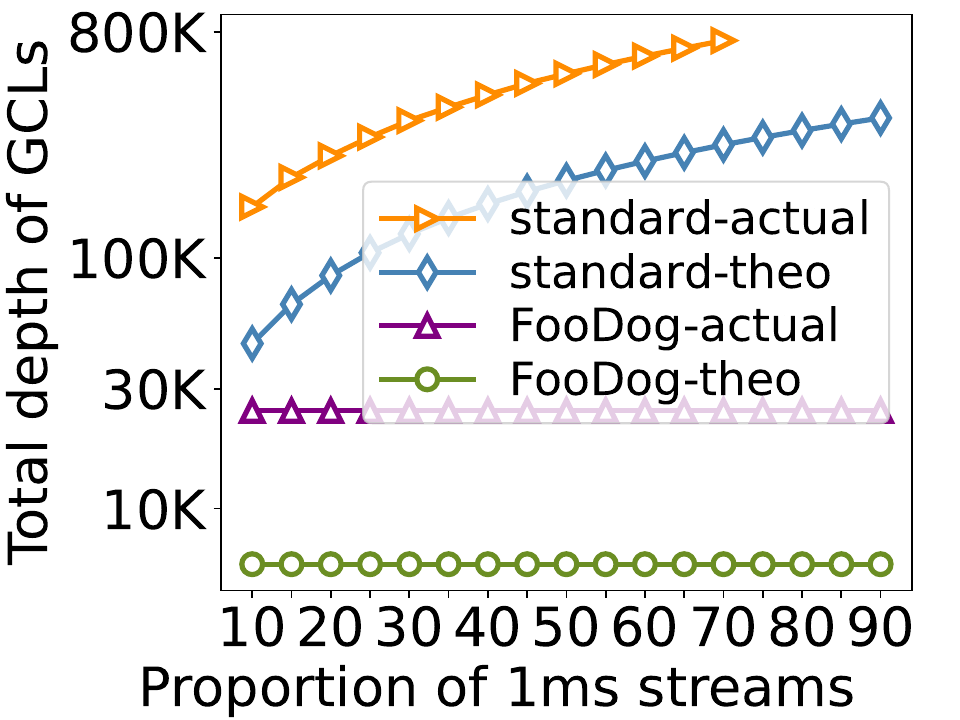}
        \caption{Total depth at different ratios of small period streams}
        \label{fig: depth}
    \end{minipage}
\end{figure}

\subsection{Memory Usage}\label{sec: memory}

The memory consumption is evaluated under a typical aerospace use case scenario. The parameters, such as the number of streams and stream periods, are selected based on a real-world use case\cite{avionic_traffic-1} and our industrial implementation. 
The number of streams ranges from 100 to 500 with a step of 50. Each scale (e.g., 100) consists of streams of a period of 1 ms and a period of 100 ms, with the proportion of 1 ms streams ranging from 10\% to 90\%. As a result, we have a total of 153 test cases, combining different numbers of streams and stream proportions ($9\times17$).
We utilize Intel Arria 10 FPGA for a comprehensive memory usage test.

\textbf{Ultra-low memory consumption.}
Figures~\ref{fig: theo memo of FooDog} and \ref{fig: exp memo of FooDog} display the theoretical and actual memory resource consumption of FooDog. The theoretical memory resource consumption is calculated using Equation~\ref{eq: FooDog usage}. On the other hand, the actual memory resource consumption is obtained from the FPGA synthesis results.
Across all test cases in the aerospace use case, the highest memory resource consumption is observed in the test where a single switch supports 500 TS streams, with 90\% of them having a period of 1ms. In this particular test case, the theoretical memory resource consumption does not exceed 0.2 Mb, while the actual resource consumption does not exceed 0.39 Mb.
Notably, the actual resource consumption only accounts for a mere 1.2\% of the total block RAMs available on the selected FPGA. This signifies that FooDog utilizes an exceptionally low amount of memory resources, showcasing its efficiency in memory utilization.

Figures~\ref{fig: theo memo of FooDog} and \ref{fig: exp memo of FooDog} demonstrate that the actual memory resource consumption is slightly larger than the theoretical consumption. This discrepancy arises from the waste of resources when concatenating basic block RAMs, such as M20K in the chosen FPGA, into larger memory arrays. During FPGA synthesis, the basic block RAMs are configured to the necessary depth and width to meet the required memory resource size. However, in some cases, the required width does not align with the supported width configuration of the block RAM. Consequently, when concatenating RAM blocks, there can be potential waste, resulting in higher actual memory resource consumption compared to the theoretical consumption.

In Figures~\ref{fig: exp memo of FooDog}, the actual memory resource consumption remains constant across all test cases. This is due to a similar reason as RAM concatenation. When the number of streams or the proportion of 1 ms stream periods is relatively small, the theoretical resource consumption is low. However, block RAM concatenation requires more memory than the theoretical resource consumption due to limitations in width and depth configuration. Specifically, 0.39 Mb memory is necessary. Even with 500 streams and a 90\% proportion of 1 ms stream periods, 0.39 Mb of memory is sufficient. This explains why a flat plane is observed in Figures~\ref{fig: exp memo of FooDog}.

\textbf{Resource reduction.}
FooDog has achieved a significant reduction in resource consumption compared to the standard PSFP design. The heatmap in Figure~\ref{fig: percentage decrease} shows the percentage of resource reduction that FooDog achieved compared to the standard PSFP design under all test cases. Across all test cases, FooDog consistently reduced resource consumption by an impressive 68\% to 98.7\%.

\textbf{Factors affecting FooDog's memory resource usage.}
The memory resource consumption of FooDog exhibits positive correlations with the number of streams and ports and is independent of the total frames of streams within the network cycle period. Figures~\ref{fig: theo memo of FooDog} and~\ref{fig: exp memo of FooDog} illustrate that FooDog's memory resource consumption is not affected by the total frames of streams within the network cycle period because in a stream scale (e.g., 100), with the proportion of streams with a period of 1ms increases, the memory usage of FooDog remains the same.
FooDog's memory resource consumption is related to the number of streams. 
In actual memory resource consumption, this linear correlation is disrupted. The disruption of this correlation in certain configurations is attributed to the concatenation of RAM blocks, as mentioned earlier. 

The memory resource consumption of FooDog demonstrates positive correlations with the number of streams and ports. However, it is independent of the total frames of streams within the network cycle period. This can be observed in Figures~\ref{fig: theo memo of FooDog} and~\ref{fig: exp memo of FooDog}, where the memory consumption remains constant regardless of the total frames of streams within the network cycle period. Because when the proportion of streams with a 1ms period increases, the memory usage of FooDog remains unchanged.

In the test case with 500 streams, 90\% of which have a period of 1ms, Figure~\ref{fig: memo of FooDog under different port} demonstrates the theoretical and actual memory resource consumption of FooDog for various numbers of ports. The figure shows that as the number of ports increases, the memory resource consumption also increases linearly.
However, even when the number of ports is increased to 16, the actual resource consumption remains below 2Mb. This is due to the memory-efficient design of our Period-wise GCL and Stream-wise GCL, which results in low resource consumption per individual port.

The memory resource consumption of PSFP is significantly influenced by the number of stream occurrences within the network cycle period, particularly when the stream periods become irregular. This irregularity leads to an increase in the total number of GCL entries due to the varying number of frames within the network cycle period. When stream periods are irregular, especially when they are coprime, resulting in a larger network cycle period, the total number of standard PSFP GCL entries increases.

Figure~\ref{fig: depth} depicts the variation in total GCL entries for both the standard PSFP design and FooDog, considering a scenario with 500 streams and an increasing proportion of streams with smaller periods. The results demonstrate that as the proportion of streams with smaller periods increases, the total number of GCL entries in the standard PSFP design also increases. 
The depth of the standard PSFP design no longer has data when the proportion exceeds 80\%.
This is because, at this stage, the resource utilization has already surpassed the total resources available on the board.
In contrast, FooDog maintains a constant number of GCL entries regardless of the proportion of streams with smaller periods by recording the planned arrival time window of the first frame of each stream.
By employing this approach, FooDog effectively overcomes the challenge posed by irregular stream periods while ensuring consistent memory resource consumption.

\subsection{Determinism}\label{sec: determinism}

\begin{figure}[htbp]
    \centering
    \begin{minipage}[t]{\linewidth}
        \centering
        \includegraphics[width=\linewidth]{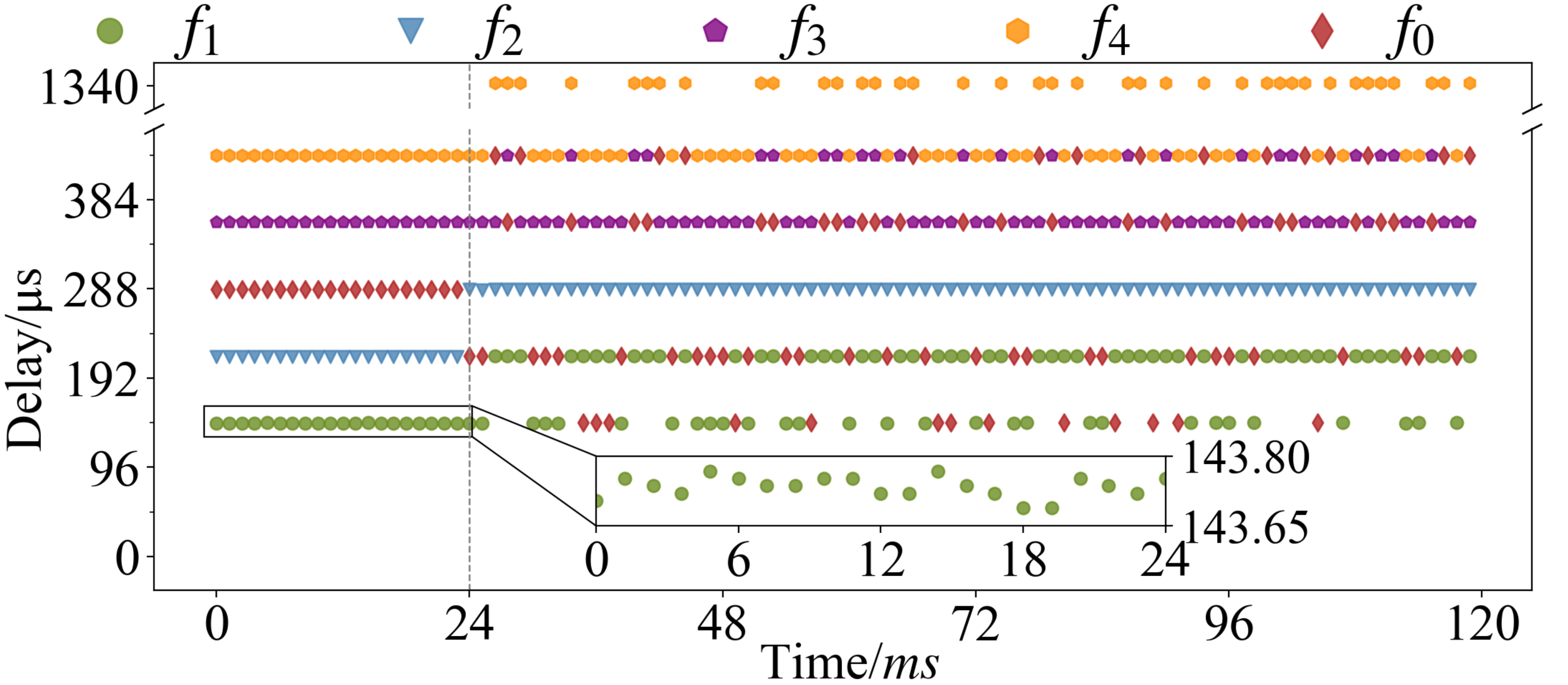}
        \caption*{(a) End-to-end delay without FooDog}
        \label{ex-fig: delay without FooDog}
    \end{minipage}
    \hfill
    \begin{minipage}[t]{\linewidth}
        \centering
        \includegraphics[width=\linewidth]{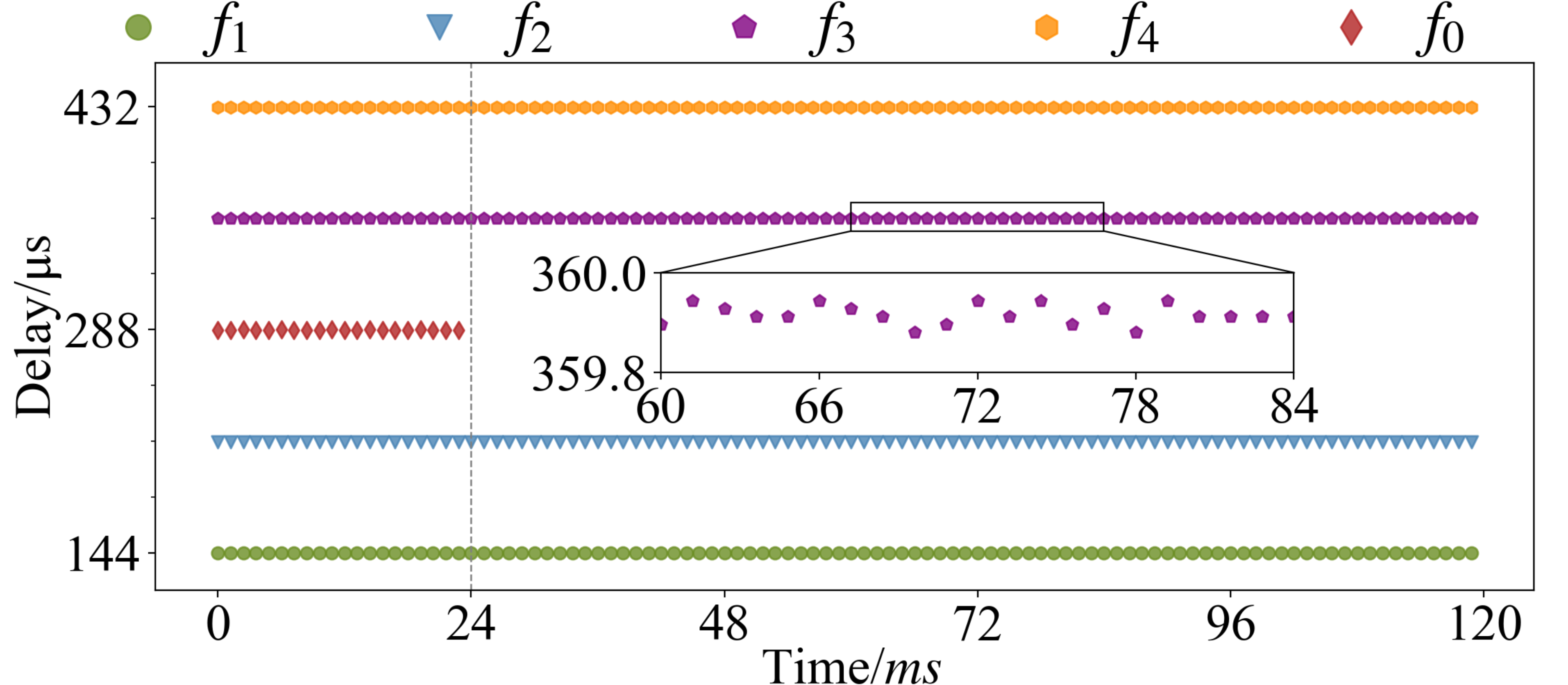}
        \caption*{(b) End-to-end delay with FooDog}
        \label{fig: delay with FooDog}
    \end{minipage}
    \\
    \begin{minipage}[t]{0.45\linewidth}
        \centering
        \includegraphics[width=\linewidth]{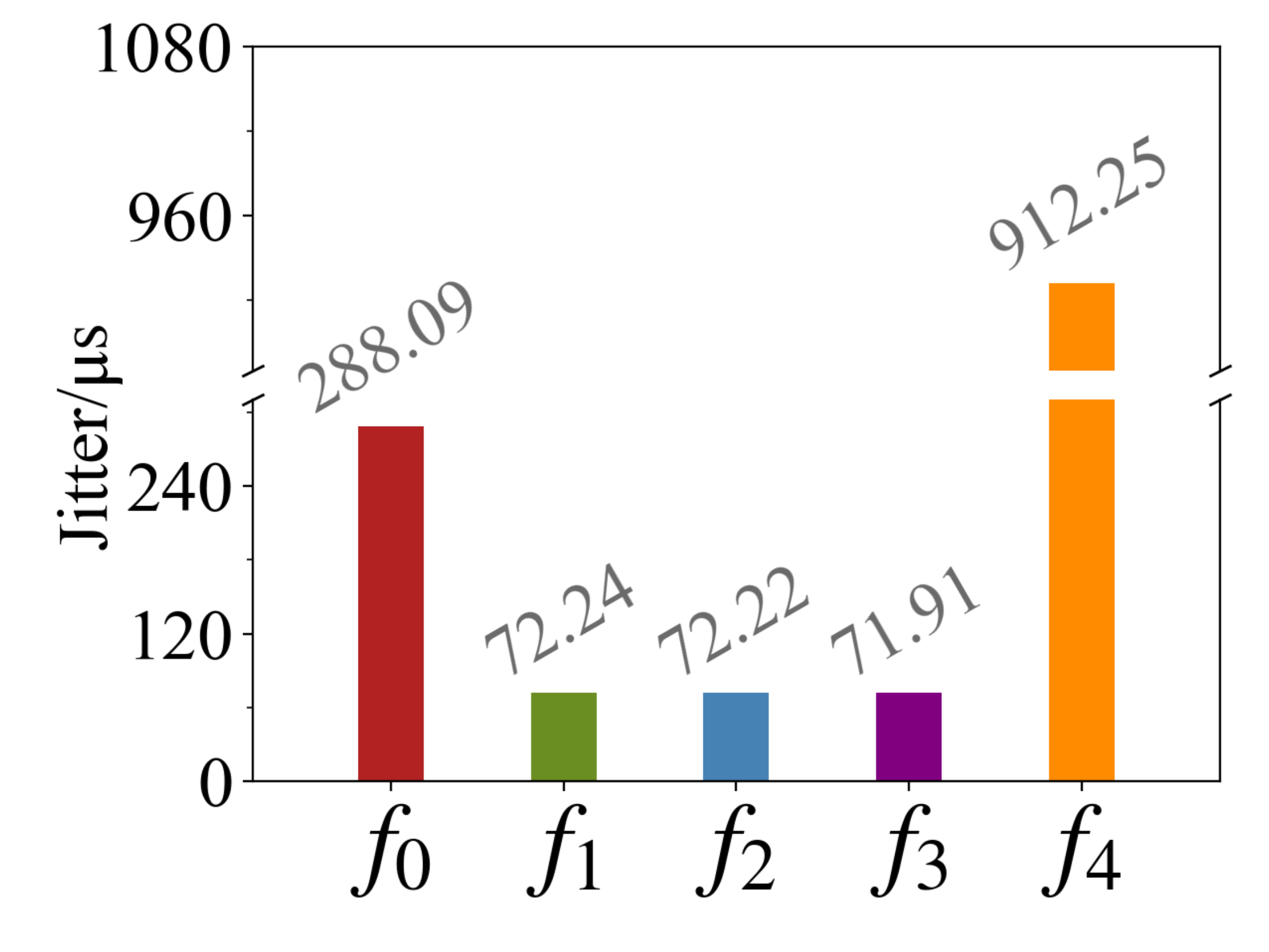}
        \caption*{(c) Jitter without FooDog}
        \label{ex-fig: jitter without FooDog}
    \end{minipage}
    \hfill
    \begin{minipage}[t]{0.45\linewidth}
        \centering
        \includegraphics[width=\linewidth]{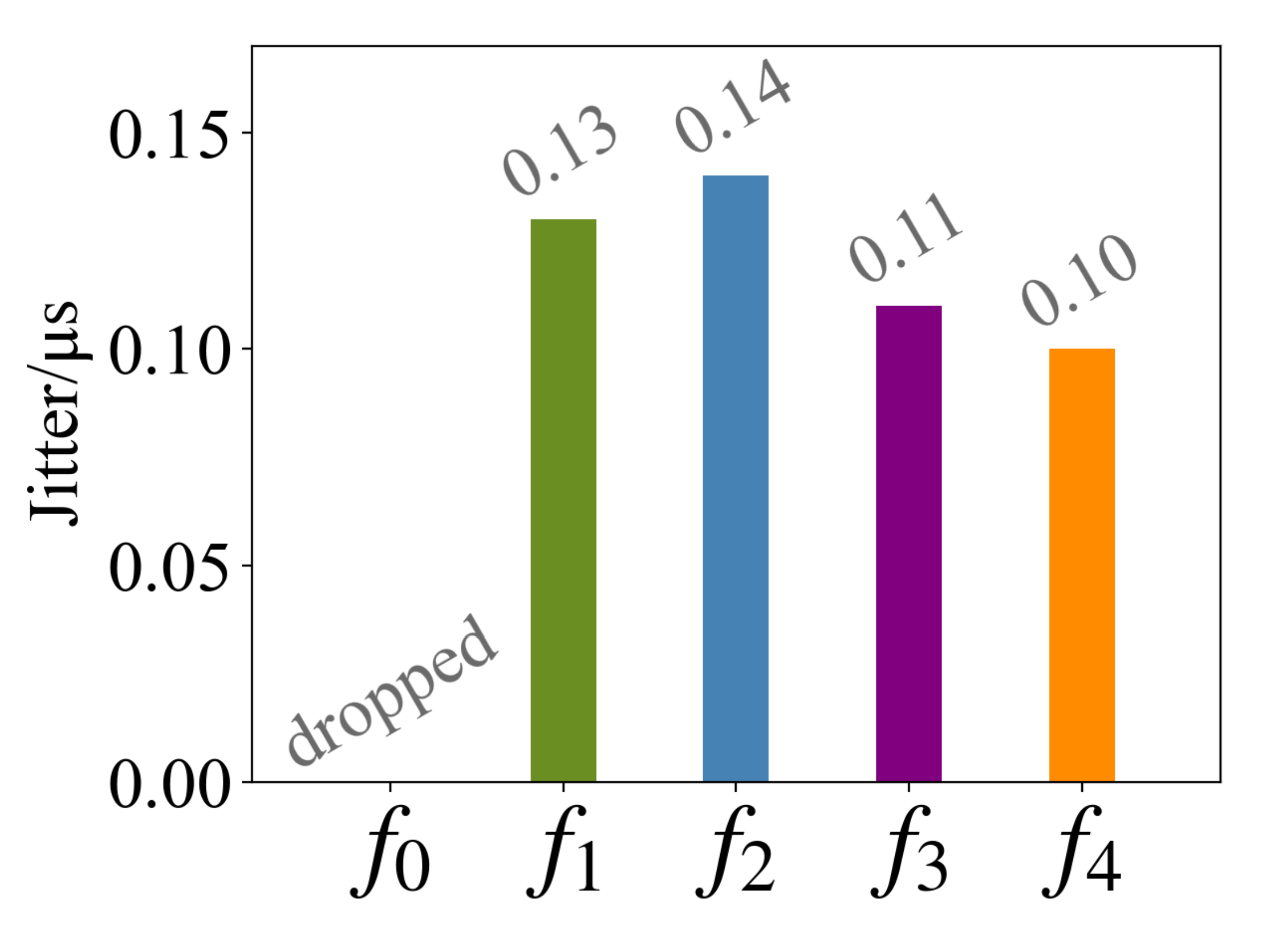}
        \caption*{(d) Jitter with FooDog}
        \label{ex-fig: jitter with FooDog}
    \end{minipage}
    \caption{Delay and jitter in avionics topology}
    \label{fig: delay and jitter}
\end{figure}

To verify the determinism offered by FooDog in real-world systems, this paper implements a prototype of FooDog using an open-source four-port TSN switch\footnote{The project name is omitted for blind review.} on Xilinx Kintex-7 XC7K480T FPGA boards. The FooDog prototype could support eight different stream periods, enabling each port to enforce policing on up to 500 streams.
The logical resources utilized by the prototype, specifically the Look-up Tables (LUTs), amount to 3896. This resource consumption represents only 0.81\% of the total LUTs available on the FPGA board.

The determinism of FooDog is evaluated based on the real-world implementations.
We have conducted an experiment using the \textit{aerospace topology}~\cite{plane_topo} as depicted in Figure~\ref{fig: aerospace}.
Wherein all TSN switches are built upon an open-source TSN project.
Period-wise GCLs and Stream-wise GCLs in TSN switches are configured using the FooDog algorithm.
There are five streams, $f_0-f_4$, which are all normal before the 24th ms.
The periods of five streams are 1ms, and the frame size is 100Bytes.
We chose to use only five streams and conduct the experiment on a small scale to specifically focus on evaluating the impact of abnormal TS streams on normal TS streams and the correctness of FooDog.
The background streams are excluded from consideration in this experiment because our primary focus lies in evaluating the influence of abnormal TS streams on normal TS streams, and the interference from the background stream is excluded by a well-established solution known as the guard band, which effectively mitigates their impact on TS streams\cite{Qbv,Fenglin}.
The end-to-end delay of the five streams without FooDog and with FooDog is collected and shown in Figure~\ref{fig: delay and jitter}\textcolor{urlblue}{(a)} and \ref{fig: delay and jitter}\textcolor{urlblue}{(b)}.

In Figure~\ref{fig: delay and jitter}\textcolor{urlblue}{(a)}, we observe that everything goes well when all streams are normal. The horizontal axis represents time, while the
vertical axis represents end-to-end delay. Between 0-24 ms, the end-to-end delay and jitter of all streams meet expectations. The zoom-in plot of $f_1$ shows the jitter is smaller than 150 nanoseconds(ns), which is sub-microsecond-level.
On the other hand, we can see that the abnormal stream would disturb the whole scheduling of TAS.
After the 24th ms, the delay of $f_1-f_4$ deteriorates immediately because the timing error of $f_0$ spreads across the whole network. Thus, streams $f_1-f_4$ experience a sharp increase in the jitter.

Figure\ref{fig: delay and jitter}\textcolor{urlblue}{(b)} illustrates the end-to-end delay of $f_0-f_4$ after enabling FooDog.
The delay of $f_0-f_1$ remains stable from 0 to the 24th ms, and after the 24th ms, FooDog discards $f_0$ due to timing errors. The normal streams $f_1-f_4$ remain unaffected by $f_0$ thanks to FooDog. Taking $f_1$ as an example, the zoom-in plot reveals only a slight variation of 200ns occurring after the 24th ms.

The comparison of the jitter of $f_0-f_4$ within the range of 0-120 ms in Figure~\ref{fig: delay and jitter}\textcolor{urlblue}{(c)} and Figure~\ref{fig: delay and jitter}\textcolor{urlblue}{(d)} is conducted to demonstrate the determinism ensured by FooDog. 
The result reveals that without employing FooDog, the jitter of $f_0-f_4$ can reach up to tens of microseconds, with even $f_4$ experiencing a jitter close to 1ms. 
Consequently, real-time requirements for $f_0-f_4$ are severely violated. 
The jitter of $f_1-f_4$ is below 150ns when employing FooDog.

\subsection{Solution Quality}\label{sec: solution quality}
The solution quality of the FooDog algorithm and the Comp algorithm is evaluated in terms of their runtime, schedulability, and end-to-end delay.

\begin{figure}[htbp]
    \centering
    \includegraphics[width=\linewidth]{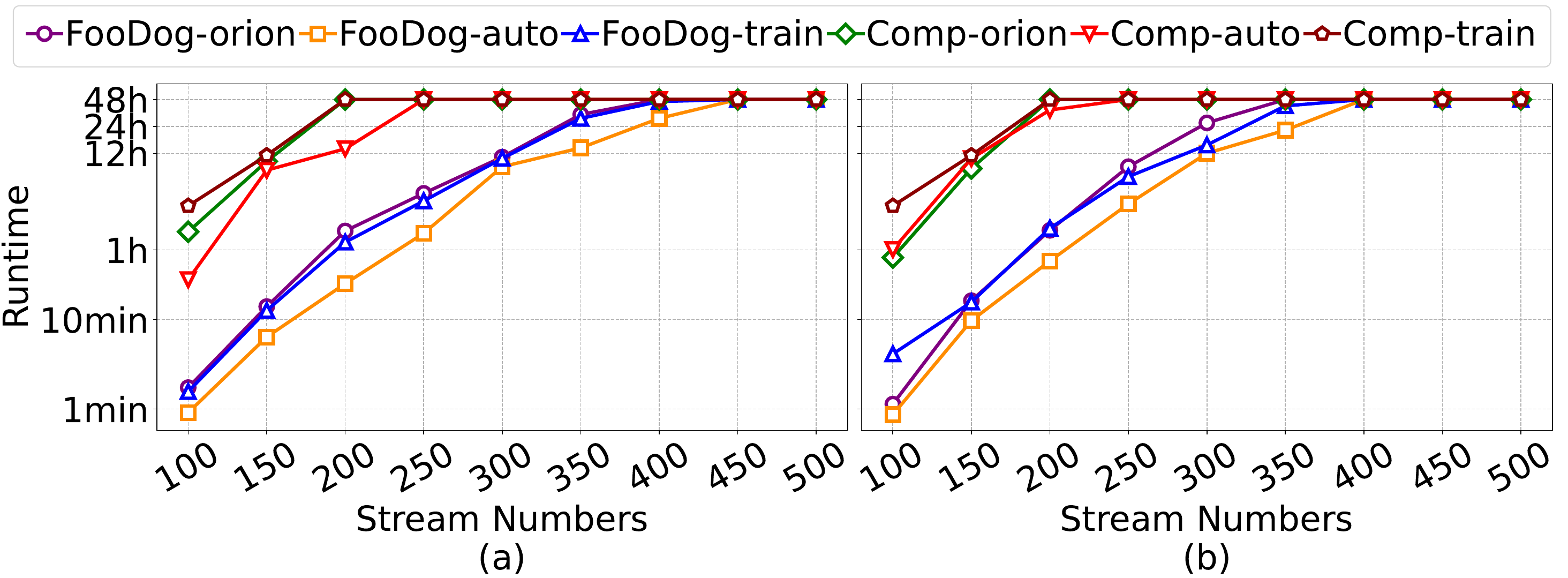}
    \caption{Runtime of FooDog algorithm and Comp algorithm: (a) streams with a period of 1ms accounts for 10\% of the total streams; (b) streams with a period of 1ms accounts for 90\% of the total streams. The y-axis is in log scale.}
    \label{ex-fig: runtime}
\end{figure}

\textbf{Runtime} experiments were conducted with a total of 9*2 configurations, varying the number of streams from 100 to 500 in increments of 50. For each stream scale, two stream configurations were tested: one with 10\% of the total streams for 1ms period and 90\% for 100ms period, and the other with 90\% for 1ms period and 10\% for 100ms period. We conducted 20 repeated experiments for each configuration and computed the average runtime. The experimental results are illustrated in Figure~\ref{ex-fig: runtime}.
As the solving tool employed by the algorithm was SMT, which exhibits high time complexity for complex problems, a timeout of 48 hours was set for algorithm execution. The experiment would be terminated if the SMT solver failed to produce a result within this deadline.
Setting a 48-hour timeout for offline algorithms provides a considerable degree of flexibility, as surpassing this threshold would likely be unfeasible in most practical situations. For instance, in the realm of smart manufacturing factories, where the duration required for executing upgrades and reconfiguration to production lines may not be allowed to exceed a full day\cite{smart}, it is crucial to consider the potential cost implications of longer upgrade times. Hence, establishing a 48-hour time limit is a justifiable choice, considering both operational constraints and economic considerations.

\begin{figure}[htbp]
    \centering
    \includegraphics[width=\linewidth]{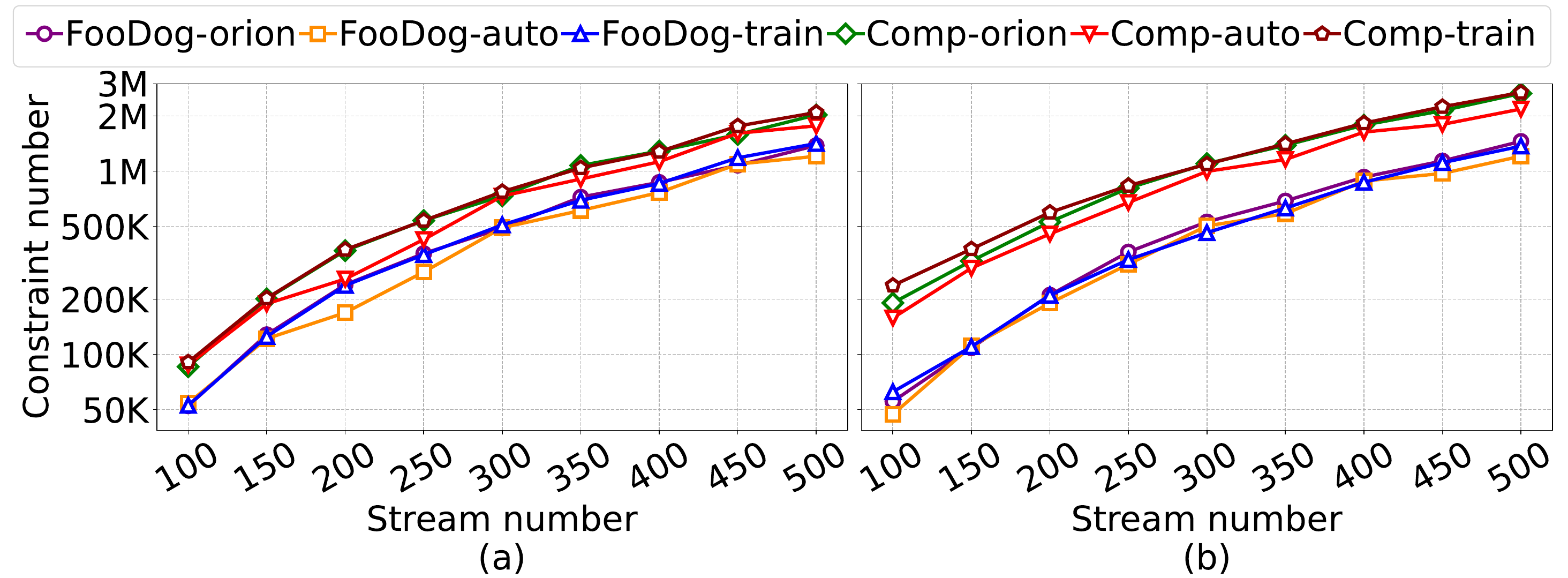}
    \caption{Constraint numbers of FooDog algorithm and Comp algorithm: (a) streams with a period of 1ms accounts for 10\% of the total streams; (b) streams with a period of 1ms accounts for 90\% of the total streams. The y-axis is in log scale.}
    \label{ex-fig: constraint number}
\end{figure}

\begin{figure}[htbp]
    \centering
    \includegraphics[width=\linewidth]{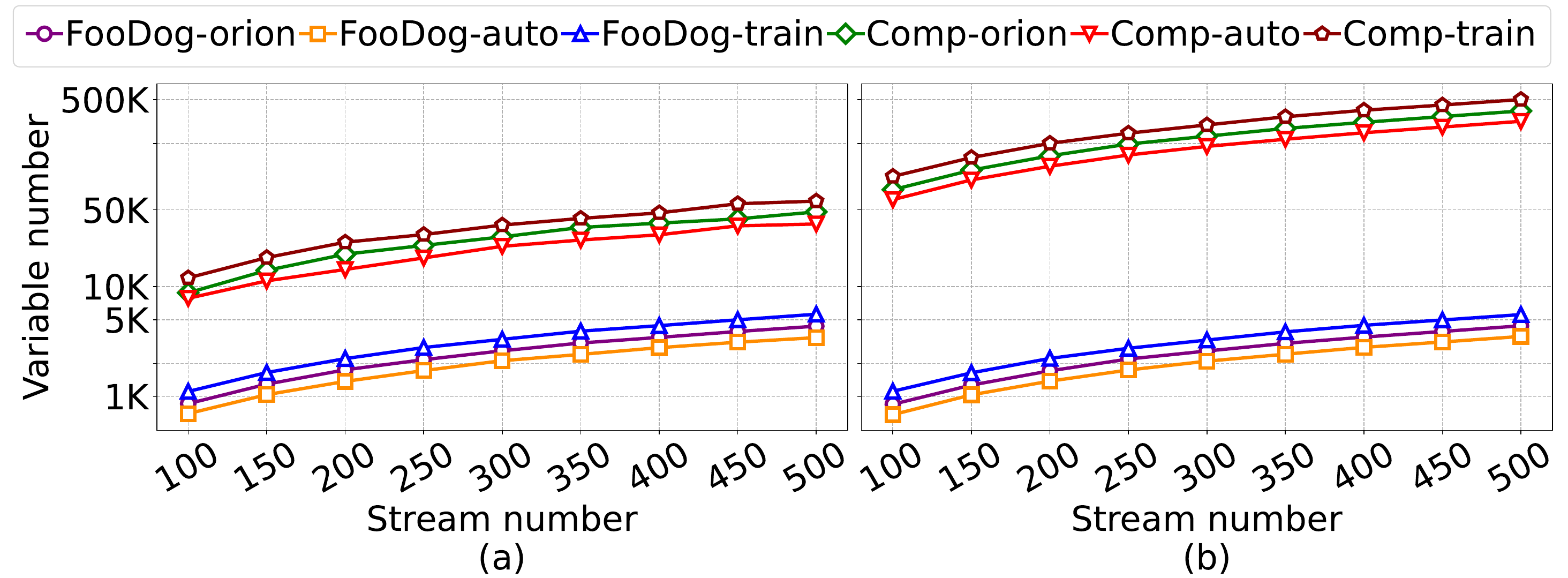}
    \caption{Variable numbers of FooDog algorithm and Comp algorithm: (a) streams with a period of 1ms accounts for 10\% of the total streams; (b) streams with a period of 1ms accounts for 90\% of the total streams. The y-axis is in log scale.}
    \label{ex-fig: variable number}
\end{figure}

As the number of streams increases, both FooDog and Comp algorithms experience exponential growth in runtime. However, the Comp algorithm consistently reaches its runtime limit faster than the FooDog algorithm due to its higher number of constraints and unknown variables, regardless of network topology. This is shown in Figures~\ref{ex-fig: constraint number} and~\ref{ex-fig: variable number}. It is unlikely that there will be significant differences in runtime trends for different constraint-solving methodologies since the number of constraints remains independent of the algorithm used. Therefore, regardless of the constraint-solving methodologies, the computational overhead can be minimized by solely addressing the determination of the transmission time for the first frame of streams within the network cycle period.

In addition, the results shown in Figure~\ref{ex-fig: runtime} indicate that the exponential increase in runtime observed in both algorithms is caused by the growing number of constraints associated with larger stream numbers. The runtime findings demonstrate a positive relationship between the proportion of 1ms periodic streams and the overall runtime. As the number of 1ms periodic streams increases relative to the total number of streams, the runtime also increases accordingly. This is due to the larger number of unknown variables and constraints that need to be solved during the computation process.

It should be noted that the prolonged runtime of the FooDog algorithm is attributed to the utilization of an SMT solver. While adopting a heuristic solver may offer faster execution, we deliberately chose to employ an SMT solver for experimental purposes due to its inherent stability. It is essential to emphasize that our objective is not focused on optimizing the algorithm's runtime but rather on conducting comprehensive experiments to compare the solution quality between algorithms with and without FooDog constraints.

\begin{figure}[htbp]
    \centering
    \includegraphics[width=\linewidth]{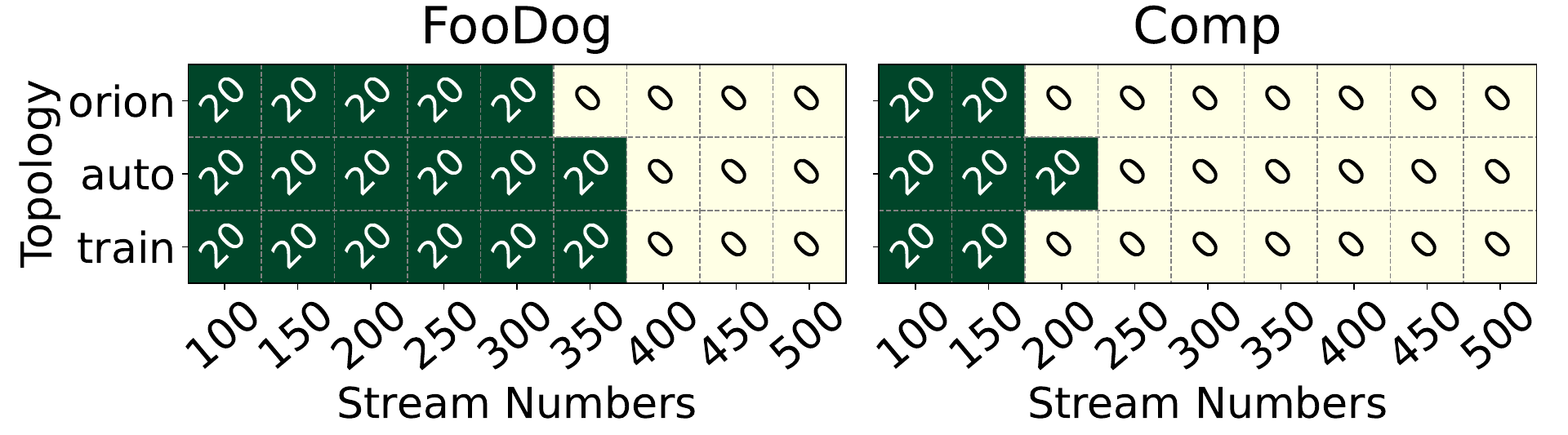}
    \caption{Scheduablity of FooDog algorithm and Comp algorithm. 20 means that all 20 repeated experiments has found a solution within 48 hours, and 0 represents that none of the 20 repeated experiments has found a solution within 48 hours.}
    \label{ex-fig: schedulablity}
\end{figure}
\begin{figure}[htbp]
    \centering
    \begin{minipage}{0.232\textwidth}
        \centering
        \includegraphics[width=\linewidth]{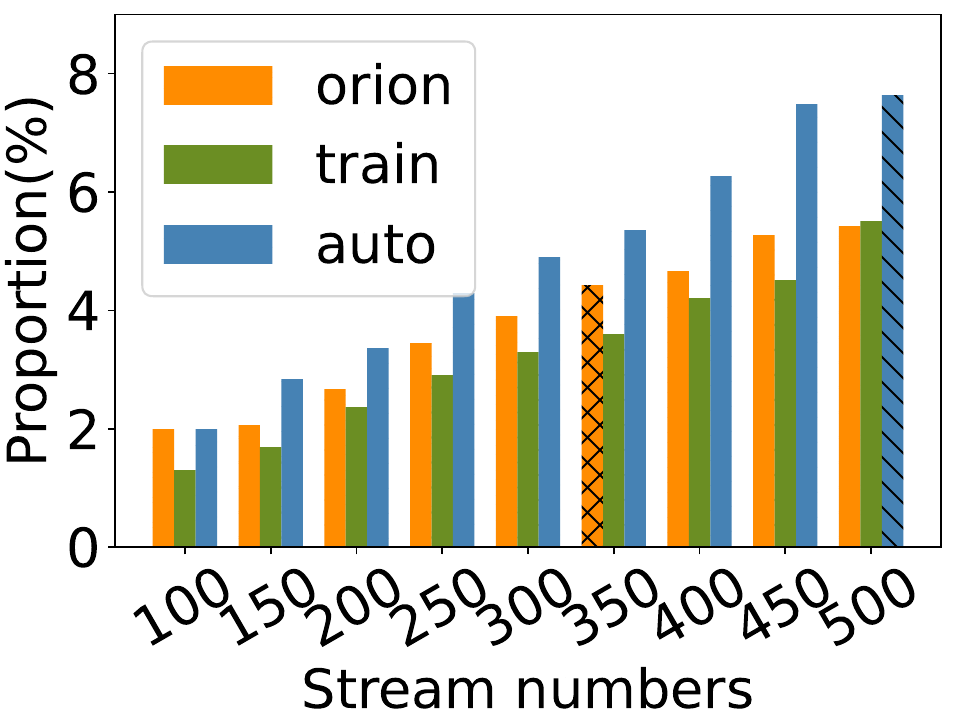}
        \caption{Proportion of maximum bandwidth usage to link speed in all 20 repeated experiments under different stream numbers}
        \label{ex-fig: bandwidth}
    \end{minipage}
    \hfill
    \begin{minipage}{0.232\textwidth}
        \centering
        \includegraphics[width=\linewidth]{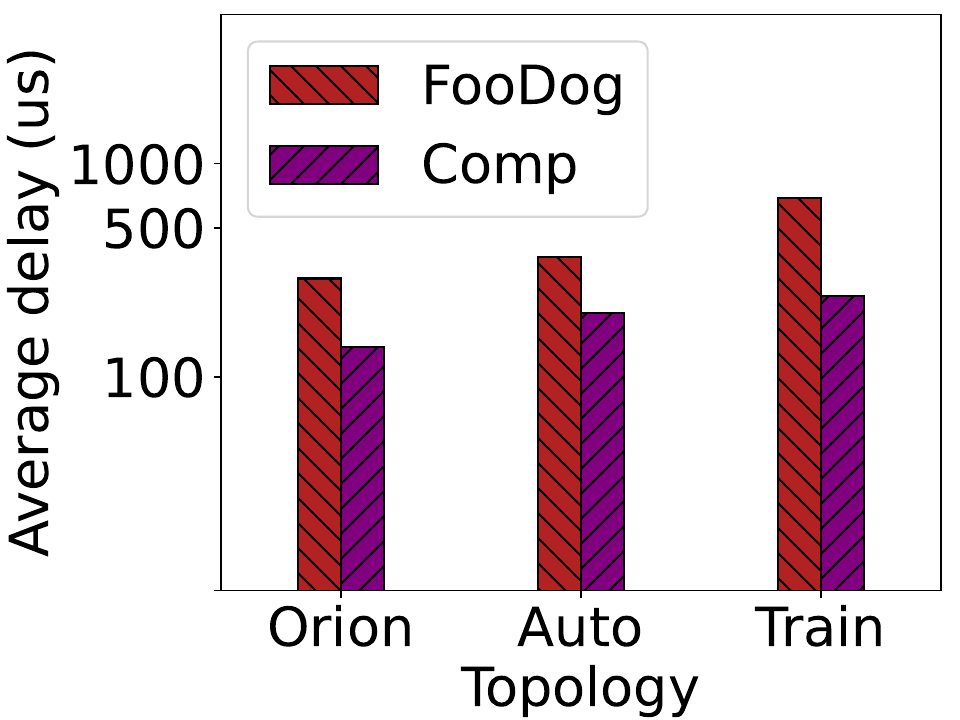}
        \caption{Average end-to-end delay calculated by FooDog and Comp algorithms under 150 streams, with 90\% 1ms-period steams}
        \label{ex-fig: end-to-end delay}
    \end{minipage}

\end{figure}

\textbf{Schedulability.}
In theory, the schedulability of the FooDog algorithm may be influenced to some extent. 
In this study, schedulability refers to the proportion of successful experiments out of the total number of repeated experiments ($M$ out of $N$) that yielded feasible solutions.
In this work, schedulability is defined as the proportion of $M$ out of $N$ repeated experiments that yielded feasible solutions. 
The results shown in Figure~\ref{ex-fig: schedulablity} demonstrate that the FooDog algorithm found a feasible solution for all 20 repeated experiments and thus achieved 100\% schedulability for all configurations within a 48-hour time limit. This indicates that the FooDog constraints did not have a significant impact in the experimental setting. One possible explanation is that the ratio of TS streams to the total bandwidth was relatively small (as shown in Figure~\ref{ex-fig: bandwidth}), allowing for a sufficiently large solution space for the problem at hand. Even with a smaller solution space, the FooDog algorithm still had enough space for feasible solutions. On the other hand, the Comp algorithm did not exhibit superior schedulability in the experimental scenario, mainly due to its excessively long solving time. In scenarios involving higher volumes of streams, it is not practical to observe the superiority of the Comp algorithm in terms of schedulability within a 48-hour deadline.

\textbf{End-to-end delay.}
Due to the stricter limitations on queue resource usage imposed by the FooDog algorithm, there is a possibility that it may result in larger end-to-end delays for TS streams. This is because these streams may have to wait longer at either the source or within switch queues to obtain the necessary resources. To assess the impact of this factor, we conducted experiments to compare the average end-to-end delay of streams generated by the FooDog and Comp algorithms. Both algorithms were given a time limit of 48 hours to find a solution. The experiments involved 150 streams, with 90\% of them having a period of 1 ms. The experimental results, depicted in Figure~\ref{ex-fig: end-to-end delay}, indicate that there were no significant differences in the quantity or order of magnitude of the average end-to-end delay observed for streams computed by the FooDog algorithm. Considering that TSN streams typically operate within the millisecond range, such an increase in end-to-end delay is generally considered acceptable.

\section{Conclusion}\label{sec: conclusion}
Time-Sensitive Networking (TSN) is a promising real-time Ethernet technology that has replaced standard Ethernet in many scenarios. TSN achieves determinism for time-sensitive (TS) traffic through Time-Aware Shaping (TAS) and global high-precision time synchronization. Each TSN switch utilizes a synchronized timetable called the TAS Gate Control List (GCL) to schedule frames on each port. The core of TAS involves an offline traffic planning algorithm that allocates exclusive time intervals for each frame of each stream, ensuring deterministic transmission. However, abnormal traffic can disrupt the ideal planning, leading to jitters and frame drops. To address this issue, Per-Stream Filtering and Policing (PSFP) is introduced to filter out abnormal traffic. This work identifies the heavy memory consumption introduced by PSFP and proposes an efficient solution called FooDog that reduces memory usage by 96\% while maintaining sub-microsecond-level performance. FooDog shows promising results in aerospace use cases and can benefit other TSN scenarios in the future.

\clearpage
\balance
\bibliographystyle{abbrv}
\bibliography{ref}

\clearpage
\appendix
\section{Constraints for Comp algorithm}\label{appendix: constraints}

The algorithm that calculates the transmission time of each stream instance uses a system model that differs from the algorithm for FooDog. Therefore, the system model and the constraints used by the comparison algorithm is described  here.

\textbf{System model.}
The network is represented using $\mathcal{G} = \{\mathcal{V}, \mathcal{E}\}$, where $\mathcal{V}$ is a set of vertices including switches and end systems, and $\mathcal{E}$ is a set of directed edges.
$(a,b)$ denotes a network link from vertex $a$ to $b$, $(a,b)\in \mathcal{E}$.
$(a,b)$ consists of four-tuples, $<BD^{(a,b)}, Q^{(a,b)}, MaxDly^{(a,b)}, MinDly^{(a,b)}, t^{(a,b)}>$, which includes the bandwidth, available transmission queues for TS streams, maximum link delay, and planning granularity respectively.
The planning granularity refers to the minimum size of the time unit used for scheduling.
The synchronization precision is denoted as $\delta$.
$\mathcal{F}$ is a set of TS streams. $f_i$ is the $i$-th TS stream,  $f_i\in \mathcal{F}, i\in [0, N-1]$, where $N$ is the total number of TS streams.
$f_i$ consists of five-tuples $<\mathcal{R}_i, C_i, T_i, L_i, J_i>$, which includes route path, frame size in bytes, period, maximum allowed end-to-end delay, and maximum allowed end-to-end jitter, respectively.
Note $T_i$ is the interval between two consecutive frames sent by the source device and is a constant value for a TS stream.
$T_i$, $L_i$ and $J_i$ are in the unit of the planning granularity.
The $f_{i,j}^{(a,b)}$ is the $j$-th frame of $f_i$ at link $(a,b)$ in a network cycle period.
The network cycle period, denoted by $T$, is the LCM of periods of all TS streams.
$\mathcal{F}^{(a,b)}$ is the set of the frames that are routed through link $(a,b)$.
Each frame $f_{i,j}^{(a,b)}$ is attached with two unknown variables that need to be solved, which are $\omega_{i}^{(a,b)}$ and $\rho_{i}^{(a,b)}$.
$\omega_{i,j}^{(a,b)}$ is the transmission time that frame $f_{i,j}^{(a,b)}$ is scheduled at link $(a,b)$, which is aligned with the planning granularity.
$\rho_{i,j}^{(a,b)}$ is the transmission queue to buffer frame $f_{i,j}^{(a,b)}$ at link $(a,b)$.

\textbf{Core Constraints.} The problem of traffic planning in TSN is a constraint-solving problem. The constraints of the comparison algorithm are described below.

\textit{Period constraint} requires that the TS frame of a period must be scheduled during the current period, as shown in Equation~\eqref{eq: period}.
There are two reasons.
First, this constraint can reduce the buffered frames in the switch.
If the $k+1$-th frame has arrived while the $k$-th frame has not been scheduled, then the switch needs to buffer two frames.
The on-chip memory is limited, however.
Second, the constraint can reduce the search space to a reasonable range.

\begin{equation}\label{eq: period}
    \begin{aligned}
        & \forall f_i \in \mathcal{F},
        \forall (a,b) \in \mathcal{R}_i, \forall j\in [0, \frac{T}{T_i}-1]:\\
        &\left(\omega_{i,j}^{(a,b)}\geq j\times T_i\right) \land \left(\omega_{i}^{(a,b)}\textless (j+1)\times T_i\right)
    \end{aligned}
\end{equation}

\textit{Contention-free constraint} ensures that frames that are transmitted through the same physical link should not overlap in the time domain, which is the basic principal of transmission in time-triggered network including TSN. The constraint is shown in Equation~\ref{eq: link}.

\begin{equation}\label{eq: link}
    \begin{aligned}
        & \forall (a,b) \in \mathcal{E}, \forall f_i^{(a,b)}, f_j^{(a,b)} \in \mathcal{F}^{(a,b)}, i\neq j, \\
        & \forall k \in [0, \frac{LCM(T_i,T_j)}{T_i}-1],
        \forall l \in [0, \frac{LCM(T_i,T_j)}{T_j}-1]:\\
        &\omega_{i,k}^{(a,b)}\geq \omega_{j,l}^{(a,b)}+\frac{C_j}{BD^{(a,b)}} \vee \omega_{j,l}^{(a,b)}\geq \omega_{i,k}^{(a,b)} + \frac{C_i}{BD^{(a,b)}}
    \end{aligned}
\end{equation}

\textit{Sequence constraint.}
The routing of a TS frame is sequential.
The scheduling time on the upstream device must be earlier than the scheduling time on the downstream device.
Besides, the scheduling time between neighbor links $(a,x)$ and $(x,b)$ must be larger than $MaxDly^{(a,x)}$ (Equation~\eqref{eq: sequence}).
Otherwise, the frame has not arrived at the downstream device, but the device has already scheduled the frame.

\begin{equation}\label{eq: sequence}
    \begin{aligned}
        &\forall f_i \in \mathcal{F}, 
        \forall (a,x),(x,b) \in \mathcal{R}_i, \forall j\in [0, \frac{T}{T_i}-1]:\\
        & \omega_{i,j}^{(x,b)}\times t^{(x,b)} \geq \omega_i^{(a,x)}\times t^{(a,x)}+\frac{C_i}{BD^{(a,x)}} + MaxDly^{(a,x)} + \delta 
    \end{aligned}
\end{equation}

\textit{End-to-end constraint.}
TS streams require stringent determinism.
The end-to-end delay must be less than the allowed maximum end-to-end delay, as shown in Equation~\eqref{eq: deadline}.
\begin{equation}\label{eq: deadline}
    \begin{aligned}
        &\forall f_i \in \mathcal{F}, 
        \exists src(\mathcal{R}_i), dst(\mathcal{R}_i),
        \forall j\in [0, \frac{T}{T_i}-1]:\\
        & \omega_{i,j}^{src(\mathcal{R}_i)}\times t^{src(\mathcal{R}_i)} + L_i \geq 
        \omega_{i,j}^{dst(\mathcal{R}_i)}\times t^{dst(\mathcal{R}_i)}
        + \frac{C_i}{BD^{dst(\mathcal{R}_i)}} + \delta
    \end{aligned}
\end{equation}

\textit{Frame isolation constraint} avoids frame interleaving.
The transmission time for any two successive TS frames from upstream links can only be in either case: (1) After the former frame has been scheduled to the downstream link, the latter frame is scheduled at the upstream link; (2) The two frames use different queues.
The formulation of the first case is as Equation~\eqref{eq: zero-aggregation 1}.

\begin{equation}\label{eq: zero-aggregation 1}
\begin{small}
    \begin{aligned}
        &\forall (a,b)\in \mathcal{E},
        \forall f^{(a,b)}_{i}, f^{(a,b)}_{j} \in \mathcal{F}^{(a,b)},
        i \neq j,\\
        & \forall k \in [0, \frac{LCM(T_i,T_j)}{T_i}-1],
        \forall l \in [0, \frac{LCM(T_i,T_j)}{T_j}-1]:\\
        & \left(\omega_{j,l}^{(a,b)}\times t^{(a,b)} + \delta \leq \omega_{i,k}^{(x,a)}\times t^{(x,a)} + MaxDly^{(x,a)}\right) \\
        & \vee 
        \left(\omega_{i,k}^{(a,b)}\times t^{(a,b)} \geq \omega_{j,l}^{(y,a)}\times t^{(y,a)} + MaxDly^{(y,a)}\right) \\
        & \vee 
        \left(\rho_{i,k}^{(a,b)} \neq \rho_{j,l}^{(a,b)}\right)
    \end{aligned}
    \end{small}
\end{equation}

The jitter constraint is not presented here.
Since there is only a single frame in the queue, the transmission of the frame is deterministic. Ideally, the end-to-end jitter depends on the synchronization precision $\delta$.

\textit{Queue constraint.}
The number of queues in the switch is typically less than 8.
The queues that can be used by TS streams are limited (Equation~\eqref{eq: queue resource}).
\begin{equation}\label{eq: queue resource}
    \begin{aligned}
        &\forall \mathcal{F}_i \in \mathcal{F},
        \forall (a,b) \in \mathcal{R}_i,
        \forall j\in [0, \frac{T}{T_i}-1]:\\
        &(\rho_{i,j}^{(a,b)} \geq 0) \land
        (\rho_{i,j}^{(a,b)} \textless Q^{(a,b)})
    \end{aligned}
\end{equation}

\section{Generation of Period-wise GCL}\label{sec: generation of Period-wise GCL}
The planning algorithm calculates the transmission timing of frames at each egress port and determines which transmission queue to buffer the frame. 
Since the planned arrival time window of a frame at downstream PSFP gates is determined by the time when the frame is scheduled by the upstream device, so the configuration of the Period-wise GCL is described by Equation~\eqref{eq: recv window}.

The upper and lower bounds of the planned arrival time window at the downstream switch meet Equation~\eqref{eq: recv window}.
\begin{equation}\label{eq: recv window}
\begin{small}
    \begin{aligned}        
    &inf|\kappa^{(a,b)}_i| = \omega^{(a,b)}_i + MinDly^{(a,b)} - \delta, \\
    &sup|\kappa^{(a,b)}_i| = \omega^{(a,b)}_i + MaxDly^{(a,b)} + \delta
    \end{aligned}
    \end{small}
\end{equation}

It describes the upper and lower bounds of the planned arrival time at the downstream switch. $(a,b)$ is a network link from device $a$ to device $b$ and $f_i^{(a,b)}$ is a frame that is routed through $(a,b)$. 
$\omega^{(a,b)}_i$ is the time that the first bit of a frame $f_i$ is scheduled by the upstream device $a$, and $f_i$ is the first frame of the TS stream $\mathcal{F}_i$ within the network cycle period.
$MinDly^{(a,b)}$ and $MaxDly^{(a,b)}$ are the minimum and the maximum link delay of $(a,b)$, respectively.
$\delta$ is the maximum deviation of any two devices' synchronized clock, i.e., synchronization precision.
$\kappa^{(a,b)}_i$ is the planned arrival time of the first bit of $f_i$ at device $b$.
Thus $inf|\kappa^{(a,b)}_i|$ is the earliest planned arrival time and $sup|\kappa^{(a,b)}_i|$ is the latest planned arrival time, which are the \textit{updateTime} of two entries in the Period-wise GCL of stream $f_i$.

\end{document}